\pdfoutput=1
\documentclass[PAPER, atlasdraft=false, USenglish, cernpreprint, texmf, orcidlogo]{atlasdoc}
\usepackage[subcaption=true]{atlaspackage}
\usepackage{atlasbiblatex}
\usepackage{siunitx}
\usepackage{float}

\usepackage{atlasphysics}
\graphicspath{{logos/}{figures/}}

\usepackage{ANA-TOPQ-2021-14-PAPER-defs}

\AtlasTitle{Search for same-charge top-quark pair production in $pp$ collisions at $\sqrt{s}=$ 13~TeV with the ATLAS detector}

\AtlasAbstract{%

A search for the production of top-quark pairs with the same electric charge ($tt$ or $\bar{t}\bar{t}$) is presented.
The analysis uses proton--proton collision data at $\sqrt{s}=13$~TeV,
recorded by the ATLAS detector at the Large Hadron Collider,
corresponding to an integrated luminosity of 140 fb$^{-1}$.
Events with two same-charge leptons and at least two $b$-tagged jets are selected.
Neural networks are employed to define two selections sensitive to additional couplings beyond the Standard Model that would enhance the production rate of same-sign top-quark pairs.
No significant signal is observed, leading to an upper limit on the
total production cross-section of same-sign top-quark pairs of 1.6 fb
at 95~\% confidence level.
Corresponding limits on the three Wilson coefficients associated with the
${\cal O}_{tu}^{(1)}$, ${\cal O}_{Qu}^{(1)}$, and ${\cal O}_{Qu}^{(8)}$
operators in the Standard Model Effective Field Theory framework are derived.

}

\AtlasRefCode{TOPQ-2021-14}

\PreprintIdNumber{CERN-EP-2024-226}

\AtlasJournal{JHEP}
\AtlasJournalRef{JHEP 02 (2025) 084}
\AtlasDOI{10.1007/JHEP02(2025)084}

\hypersetup{pdftitle={ATLAS document},pdfauthor={The ATLAS Collaboration}}

\begin{document}

\maketitle

\tableofcontents

%
%

\section{Introduction}
\label{sec:intro}

\begin{figure}[h]
\centering
\begin{subfigure}[b]{0.4\textwidth}
\includegraphics[width=\textwidth]{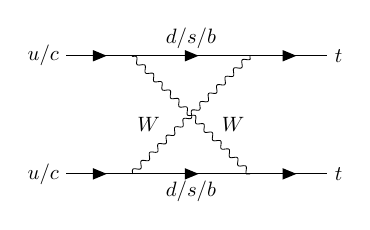}
\caption{}
\label{fig:SStopSM}
\end{subfigure}
\begin{subfigure}[b]{0.3\textwidth}
\includegraphics[width=\textwidth]{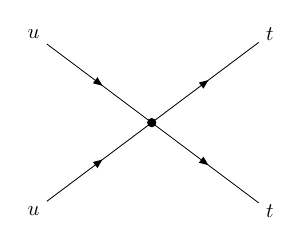}
\caption{}
\label{fig:SStopEFT}
\end{subfigure}
\caption{Lowest-order Feynman diagrams for same-sign top-quark pair production (a) in the SM and (b) as a pointlike effective-field-theory (EFT) four-fermion interaction.}
\label{fig:SStopDiagrams}
\end{figure}

The search for exclusive production of top-quark pairs with the same electric charge (same-sign, SS)\footnote{For simplicity, throughout this paper charge conjugation is implicitly assumed, therefore ``same-sign top-quark pairs'' refers to both $tt$ and $\overline{t}\overline{t}$. }
is driven by the potential to uncover phenomena beyond the Standard Model (BSM) of particle physics.
In the Standard Model (SM), SS top-quark pair production
is forbidden at leading order (LO) in perturbation theory.
This process can only occur at the lowest perturbative order through the exchange
of two $W$ bosons, as shown in Figure~\ref{fig:SStopSM}. The cross-section for this process is
calculated to be approximately $\sigma(pp \to tt)_{SM} \simeq 4 \cdot 10^{-15}$~pb
for proton--proton collisions at $\sqrt{s} = 13$~\TeV~ \cite{https://doi.org/10.48550/arxiv.1507.00020},
which is not detectable at the Large Hadron Collider (LHC).
The observation of this process at the LHC would therefore imply the existence of BSM physics processes.

Model-independent searches for BSM processes at high energies can be conducted using
the Standard Model Effective Field Theory (SMEFT). In this approach, the SM
is assumed to be a LO and low-energy approximation of a more fundamental
theory, with a new-physics energy scale $\Lambda$ substantially
larger than the SM electroweak (EW) scale.

In this paper, a search for SS top-quark pair production is presented.
The process is described, using the SMEFT framework, as a pointlike
interaction involving four up-type fermions as shown in Figure~\ref{fig:SStopEFT}.
Among the complete set of dimension-6 EFT operators expressed in the Warsaw basis~\cite{Grzadkowski_2010},
only five can be responsible for the production of this SS top-quark signal~\cite{DEGRANDE2011306} with a flavor universality assumption. The following notation is used for the operators following the conventions of Ref.~\cite{DEGRANDE2011306}:
\begin{align*}{\cal O}_{tu}^{(1)} &= [ \overline{t}_\text{R} \gamma^\mu u_\text{R} ]  [ \overline{t}_\text{R} \gamma_\mu u_\text{R} ] , \nonumber \\
{\cal O}_{Qq}^{(1)} &= [ \overline{Q}_\text{L} \gamma^\mu q_\text{L} ]  [ \overline{Q}_\text{L} \gamma_\mu q_\text{L} ] , \nonumber \\
{\cal O}_{Qq}^{(3)} &= [ \overline{Q}_\text{L} \gamma^\mu \sigma^a q_\text{L} ]  [ \overline{Q}_\text{L} \gamma_\mu \sigma^a q_\text{L} ] ,  \nonumber \\
{\cal O}_{Qu}^{(1)} &= [ \overline{Q}_\text{L} \gamma^\mu q_\text{L} ]  [ \overline{t}_\text{R} \gamma_\mu u_\text{R} ] ,  \nonumber \\
{\cal O}_{Qu}^{(8)} &= [ \overline{Q}_\text{L} \gamma^\mu T^A q_\text{L} ]  [ \overline{t}_\text{R} \gamma_\mu T^A u_\text{R} ] .  \nonumber
\end{align*}
In these expressions, $Q_\text{L}$ and $t_\text{R}$ are the left-handed doublet and right-handed singlet of the third quark generation, $q_\text{L}$ and $u_\text{R}$
are related to the first two generations, and $\sigma^a$ and $T^A$ are the generators of SU(2)$_\text{L}$ and SU(3)$_\text{C}$, respectively.
Only ${\cal O}_{tu}^{(1)}$, ${\cal O}_{Qu}^{(1)}$, and ${\cal O}_{Qu}^{(8)}$ are
considered, since ${\cal O}_{Qq}^{(1)}$ and ${\cal O}_{Qq}^{(3)}$ are already constrained
by $B_d$ mixing and dijet production measurements~\cite{UTfit:2007eik}.
The effective Lagrangian related to these operators is expressed as:
\begin{equation}{\cal L}_{\mathrm D=6}^{qq \to tt} = \frac{1}{\Lambda^2}
\left( c_{tu}^{(1)} {\cal O}_{tu}^{(1)}
+ c_{Qu}^{(1)} {\cal O}_{Qu}^{(1)}
+ c_{Qu}^{(8)} {\cal O}_{Qu}^{(8)}
\right) + h. c.
\end{equation}
where $c_{tu}^{(1)}$, $c_{Qu}^{(1)}$ and $c_{Qu}^{(8)}$ are the Wilson coefficients (WCs) related to the EFT operators.

Searches for new physics processes containing two SS leptons
have only small background contributions from known SM processes,
such as diboson production including vector boson scattering $W^{\pm}W^{\pm}$,
and $t \overline{t} W$, $t \overline{t} Z$ and $t \overline{t} H$ production
in the multilepton final state, where one or more leptons may be undetected.

Both the ATLAS and CMS collaborations have searched for
anomalous production of events containing pairs of SS leptons at the LHC,
using proton--proton $(pp)$ collision data at $\sqrt{s}=7,8$ and 13~\TeV~\cite{CMS-EXO-11-065,TOPQ-2011-19,EXOT-2013-16,CMS-SUS-16-035,EXOT-2016-16,CMS-SUS-19-008}.

A previous ATLAS analysis~\cite{EXOT-2013-16} with 20~fb$^{-1}$ of $pp$ collision data at $\sqrt{s}=8$~\TeV
set 95\% confidence level (CL) upper limits on $\sigma (pp \to tt)$ at 62, 51 and 38 fb
assuming a contact interaction for left--left, left--right and right--right chirality of the SS top quarks, respectively.
This corresponds to limits on the WCs of $|c|/\Lambda^2<0.053, 0.137$ and $0.042$~\TeV$^{-2}$~\cite{EXOT-2013-16}.
A 95\% CL limit on the $tt$ production cross-section at $\sqrt{s}=13$~\TeV
of $\sigma (pp \to tt) < 89$~fb is set by the ATLAS Collaboration
using 36~fb$^{-1}$ of $pp$ collision data, assuming a
BSM flavor-changing neutral current mediator with a mass of 1~\TeV~\cite{EXOT-2016-16}.

In this paper, a dedicated analysis is designed to specifically target the search for SS top-quark pair production with a minimized model dependence.


\section{ATLAS detector}
\label{sec:detector}

\newcommand{\AtlasCoordFootnote}{%
ATLAS uses a right-handed coordinate system with its origin at the nominal interaction point (IP)
in the centre of the detector and the \(z\)-axis along the beam pipe.
The \(x\)-axis points from the IP to the centre of the LHC ring,
and the \(y\)-axis points upwards.
Polar coordinates \((r,\phi)\) are used in the transverse plane,
\(\phi\) being the azimuthal angle around the \(z\)-axis.
The pseudorapidity is defined in terms of the polar angle \(\theta\) as \(\eta = -\ln \tan(\theta/2)\) and is equal to the rapidity
$ y = \frac{1}{2} \ln \left( \frac{E + p_z c}{E - p_z c} \right) $ in the relativistic limit.
Angular distance is measured in units of \(\Delta R \equiv \sqrt{(\Delta y)^{2} + (\Delta\phi)^{2}}\).}

The ATLAS detector~\cite{PERF-2007-01} at the LHC covers nearly the entire solid angle around the collision point.\footnote{\AtlasCoordFootnote}
It consists of an inner tracking detector surrounded by a thin superconducting solenoid, electromagnetic and hadron calorimeters,
and a muon spectrometer incorporating three large superconducting air-core toroidal magnets.

The inner-detector system (ID) is immersed in a \qty{2}{\tesla} axial magnetic field
and provides charged-particle tracking in the range of \(|\eta| < 2.5\).
The high-granularity silicon pixel detector covers the vertex region and typically provides four measurements per track,
the first hit normally being in the insertable B-layer. %
It is followed by the silicon microstrip tracker, which usually provides eight measurements per track.
These silicon detectors are complemented by the transition radiation tracker (TRT),
which enables radially extended track reconstruction up to \(|\eta| = 2.0\).
The TRT also provides electron identification information
based on the fraction of hits (typically 30 in total) above a higher energy-deposit threshold corresponding to transition radiation.

The calorimeter system covers the pseudorapidity range \(|\eta| < 4.9\).
In the region \(|\eta|< 3.2\), electromagnetic calorimetry is provided by barrel and
endcap high-granularity lead/liquid-argon (LAr) calorimeters,
with an additional thin LAr presampler covering \(|\eta| < 1.8\)
to correct for energy loss in material upstream of the calorimeters.
Hadron calorimetry is provided by the steel/scintillator-tile calorimeter,
segmented into three barrel structures with \(|\eta| < 1.7\), and two copper/LAr hadron endcap calorimeters.
The solid angle coverage is completed with forward copper/LAr and tungsten/LAr calorimeter modules
optimized for electromagnetic and hadronic energy measurements respectively.

The muon spectrometer (MS) comprises separate trigger and
high-precision tracking chambers measuring the deflection of muons in a magnetic field generated by the superconducting air-core toroidal magnets.
The field integral of the toroids ranges between \num{2.0} and \qty{6.0}{\tesla\metre}
across most of the detector.
Three layers of precision chambers, each consisting of layers of monitored drift tubes, cover the region \(|\eta| < 2.7\),
complemented by cathode-strip chambers in the forward region, where the background is highest.
The muon trigger system covers the range \(|\eta| < 2.4\) with resistive-plate chambers in the barrel, and thin-gap chambers in the endcap regions.

Interesting events are selected by the first-level trigger system implemented in custom hardware,
followed by selections made by algorithms implemented in software in the high-level trigger~\cite{TRIG-2016-01}.
The first-level trigger accepts events from the \qty{40}{\MHz} bunch crossings at a rate below \qty{100}{\kHz},
which the high-level trigger further reduces in order to record events to disk at an average rate of about \qty{1}{\kHz}.

An extensive software suite~\cite{SOFT-2022-02} is used in data simulation, in the reconstruction
and analysis of real and simulated data, in detector operations, and in the trigger and data acquisition
systems of the experiment.


\section{Data and simulated events}
\label{sec:datamcsamples}

This analysis uses $pp$ collision data at $\sqrt{s}=13~\tev$ collected with the ATLAS detector between 2015 and 2018.
After the application of data-quality requirements~\cite{DAPR-2018-01}, the data sample corresponds to an integrated luminosity of
$140~\ifb$~\cite{DAPR-2021-01}. Only events recorded under stable beam conditions and for which all detector subsystems
were known to be in a good operating condition are used.

Monte Carlo (MC) simulated events were produced to study signal and background processes.
The simulated events were processed through either a full simulation of the ATLAS detector
using \textsc{Geant4}~\cite{SOFT-2010-01,Agostinelli:2002hh}, or a fast simulation, in which the
simulation of the calorimeter response is replaced by a parameterization
of the shower shapes~\cite{ATL-SOFT-PUB-2018-002}. Both types of simulated events were processed through the
same reconstruction software used for the $pp$ collision data.

Additional inelastic $pp$ collisions from the same and neighboring bunch crossings (pileup) are
modeled using events from minimum-bias interactions simulated with \PYTHIA[8.186]~\cite{Sjostrand:2007gs}
using the \NNPDF[2.3lo] set of parton distribution functions (PDFs)~\cite{Ball:2012cx} and the A3 set of tunable parameters~\cite{ATL-PHYS-PUB-2016-017}, referred to as ``tune''.
They were overlaid onto the simulated hard-scatter events according to the luminosity profile of the recorded data.
The number of $pp$ interactions per bunch crossing in the data used here ranges from about 8 to 70, with an average of 34.

Unless otherwise specified, events generated with \MGNLO~\cite{Alwall:2014hca} or \POWHEGBOX~\cite{powhegtt,powhegstp,powhegstp2}
were interfaced to \PYTHIA[8]~\cite{Sjostrand:2014zea} and used the A14 tune ~\cite{ATL-PHYS-PUB-2014-021} with the \NNPDF[2.3lo] PDF set to model the parton shower (PS),
underlying event (UE) and hadronization. Events interfaced to
\HERWIG[7]~\cite{Bahr:2008pv,Bellm:2015jjp} used the
\MMHT[lo] PDF set~\cite{Harland-Lang:2014zoa} and the corresponding tune provided by \HERWIG, H7.2-Default or H7-UE-MMHT\@.
Events simulated with \SHERPA~\cite{Bothmann:2019yzt} used the PS and hadronization model and the tune provided by the \SHERPA authors.
All simulated events, except those produced with the \SHERPA event generator,
used \textsc{EvtGen}~\cite{Lange:2001uf} to model the decays of heavy-flavor hadrons.
The mass of the top quark and the Higgs boson were set to 172.5~\GeV and 125~\GeV, respectively.

Samples of simulated MC events were produced for the different signal and background processes.
The generators used for the matrix element (ME) and PS simulation and the PDF configurations of all samples are summarized
in Table~\ref{tab:mcconfig}, with the samples used to estimate the systematic uncertainties that are described in Section~\ref{sec:modeling} given in parentheses.

\subsection{Signal simulation}

SS top-quark signal samples were simulated using \MGNLO v2.9.3
together with the \textsc{SMEFTsim3.0} Feynrules package ~\cite{Brivio:2017btx,Brivio:2020onw,Alloul_2014},
using the $m_W$ electroweak input scheme.
Only the three operators ${\cal O}_{tu}^{(1)}$, ${\cal O}_{Qu}^{(1)}$,
and ${\cal O}_{Qu}^{(8)}$ were considered.
Flavor universality was imposed, so that the WCs for $uutt$, $uctt$, and $cctt$ vertices
were bound to the same value.
The ME calculation for the signal processes was performed at LO in quantum chromodynamics (QCD) without additional jets using the \NNPDF[30lo]~\cite{Ball:2014uwa} PDF set.\footnote{The ME calculation is at LO and the jet multiplicity could hence be mismodeled. This was tested by applying scale variations to the jet multiplicity distribution and it was found to have a negligible impact on the results.}
Top quarks were decayed into a $W$ boson and a $b$-quark, and the $W$ bosons subsequently decayed into a charged lepton and a neutrino of the same flavor.

A nominal sample was produced, in which the WCs were set to $c_{tu}^{(1)}=0.04$, $c_{Qu}^{(1)}=0.1$ and
$c_{Qu}^{(8)}=0.1$. This set of WC values corresponds to cross-sections of $\sigma(pp \rightarrow tt) =$ 97.6~fb and $\sigma(pp \rightarrow \bar{t}\bar{t}) =$ 2.4~fb, assuming a new-physics scale $\Lambda = 1$~\TeV.
The SS top-quark signal is highly charge-asymmetric leading to a cross-section which is approximately 40 times larger for $tt$ production compared to $\bar{t}\bar{t}$ production.
Instead of generating multiple signal samples associated with different assumptions on WCs,
the \MGNLO reweighting tool~\cite{Mattelaer_2016} was exploited to obtain the necessary set of
signal predictions from the nominal sample.
An event weight for a different set of WCs is obtained as:
\begin{equation}
W_{\textrm{new}} = \frac{| M_{\textrm{new}} |^2}{| M_{\textrm{orig}} |^2}  W_{\textrm{orig}}
\end{equation}
where $W_{\textrm{orig}}$ is the original event weight and $M_{\textrm{new}}$, $M_{\textrm{orig}}$ are the
re-computed and original MEs, respectively.
The chosen ranges for the WCs, $c_{tu}^{(1)}=[0.01,0.05]$, $c_{Qu}^{(1)}=[0.05,0.15]$,
and $c_{Qu}^{(8)}=[0.10,0.30]$, imply values for the SS top pair production cross-section in
the dileptonic decay channel ranging from 1~fb to 100~fb, approximately, assuming a new-physics scale $\Lambda = 1$~\TeV.

\begin{table}[!h]
\begin{center}
\caption{\label{tab:mcconfig}
The samples used for event generation of signal and background processes.
Those used to estimate the systematic uncertainties are shown in parentheses.
The production of an electroweak boson is denoted by $V$ ($W$ or $Z/\gamma^*$).
The parton density function (PDF) shown in the table is the one used for the matrix element (ME).
The tune is the set of tuned parameters of the parton-shower (PS) generator.}

{\small
\setlength\tabcolsep{1.5pt}
\begin{tabular}{llllll}
\toprule
Process & Generator & ME order & PS & PDF (ME) & Tune  \\
\midrule
$tt/\bar{t}\bar{t}$ EFT signal \;\; & \MGNLO & LO & \PYTHIA[8] & \NNPDF[3.0lo] & A14 \\
$\ttbar W$             & \SHERPA[2.2.10] & NLO & \SHERPA &  \NNPDF[3.0nnlo] & \SHERPA default \\
& (\MGNLO)  & (NLO) & (\PYTHIA[8]) & (\NNPDF[3.0nlo]) & (A14)   \\
& (\POWHEGBOX) & (NLO) & (\PYTHIA[8]) & (\NNPDF[2.3lo]) & (A14) \\
& (\POWHEGBOX) & (NLO) & (\HERWIG[7.2]) & (\NNPDF[3.0nlo]) & (H7.2-Default)\\
$t\overline{t}Z/\gamma^{*}$  & \MGNLO      & NLO & \PYTHIA[8]    & \NNPDF[3.0nnlo] & A14\\
& (\MGNLO) & (NLO) & (\HERWIG[7.2]) & (\NNPDF[3.0nlo]) & (H7.2-Default)  \\
& (\MGNLO) & (NLO) & (\PYTHIA[8]) & (\NNPDF[3.0nlo]) & (A14 Var3c)  \\
$\ttbar ll$ & \MGNLO & NLO & \PYTHIA[8] & \NNPDF[3.0nlo] & A14   \\
\ttH                   & \POWHEGBOX  & NLO & \PYTHIA[8]\ & \NNPDF[3.0nlo] & A14 \\
& (\POWHEGBOX) & (NLO) & (\HERWIG[7.04]) &  (\NNPDF[3.0nlo]) &  (H7UE-MMHT) \\
& (\MGNLO)  & (NLO) & (\PYTHIA[8]) & (\NNPDF[3.0nlo]) & (A14)   \\
$VV$, $qqVV$, $VVV$         & \SHERPA[2.2.2]  & NLO & \SHERPA & \NNPDF[3.0nnlo] & \SHERPA default  \\
Four top      & \MGNLO & NLO & \PYTHIA[8] & \NNPDF[3.1nlo] & A14 \\
$\ttbar$               & \POWHEGBOX & NLO & \PYTHIA[8] & \NNPDF[3.0nlo] & A14  \\
$s$-, $t$-channel,     & \POWHEGBOX & NLO & \PYTHIA[8] & \NNPDF[3.0nlo] & A14   \\
$Wt$ single top        & & &  \\
$Z \to l^+l^-$ (Mat Conv) & \POWHEGBOX & NLO & \PYTHIA[8] & \CT[10nlo] & AZNLO \\
$Z \to l^+l^-$+($\gamma*$) & \POWHEGBOX & NLO & \PYTHIA[8] & \CT[10nlo] & AZNLO \\
$Z \to l^+l^-$         & \SHERPA[2.2.1] & NLO & \SHERPA & \NNPDF[3.0nnlo] & \SHERPA default \\
$W+$jets               & \SHERPA[2.2.1] & NLO & \SHERPA & \NNPDF[3.0nnlo] & \SHERPA default \\
$V \gamma$			   & \SHERPA[2.2.8] & NLO & \SHERPA & \NNPDF[3.0nnlo] & \SHERPA default \\
$t (Z/\gamma^*)$, $t\bar t t$, $t \bar{t} WH$  & \MGNLO & LO & \PYTHIA[8] & \NNPDF[2.3lo] & A14  \\
$t\bar t W^+ W^-$, $t \bar{t} ZZ$, $t \bar{t} HH$ \, \, & \MGNLO & LO  & \PYTHIA[8] & \NNPDF[2.3lo] & A14  \\
$t W (Z/\gamma^*)$, $tWH$, $ tHqb$  & \MGNLO & NLO & \PYTHIA[8] & \NNPDF[3.0nlo] & A14  \\
$VH$                   & \POWHEGBOX & NLO  & \PYTHIA[8] & \NNPDF[3.0nlo] & A14 \\
\bottomrule
\end{tabular}
}
\end{center}
\end{table}

\subsection{Simulation of background events}

SM processes such as the production of \ttW, \ttZ, \ttg, \ttH, \tttt, $WW$+jets, $WZ$+jets and $ZZ$+jets
can have two leptons of the same electric charge in the final state and mimic the signal signature.

The \ttW production is expected to constitute the largest fraction of the background in the signal-dominated regions. The used simulation setup for \ttW production follows Ref.~\cite{TOPQ-2019-30}.
It was simulated using \SHERPA[2.2.10].
Both the factorization and renormalization scales, $\mu_\text{F}$ and $\mu_\text{R}$, were set to $H_T/2$, where $H_T$ is the sum of
the transverse masses $m_{T} = \sqrt{m^2 + p_{T}^2 }$ of all particles generated in the ME calculation.
The sample was simulated at next-to-leading order (NLO) in QCD accuracy for MEs for up to one jet and LO accuracy for up to two additional jets.
The additional partons were matched and merged with the \Sherpa PS based on Catani--Seymour dipole
factorization~\cite{Gleisberg:2008fv,Schumann:2007mg}, using the \textsc{MEPS@NLO} prescription~\cite{Hoeche:2011fd,Hoeche:2012yf,Catani:2001cc,Hoeche:2009rj}
with a CKKW merging scale of 30~\gev.
The virtual QCD corrections for MEs at NLO accuracy were provided by the \textsc{OpenLoops}~2 library~\cite{Buccioni:2019sur,Cascioli:2011va,Denner:2016kdg}.
Samples were simulated using the \NNPDF[3.0nnlo]~\cite{Ball:2014uwa} PDF set.
The LO EW contributions were obtained from a dedicated sample simulated with \Sherpa~2.2.10 and were combined
with the NLO QCD sample described above.
Three alternative \ttW samples are used in the estimation of systematic uncertainties and are listed in Table~\ref{tab:mcconfig}.
The events were normalized using a cross-section of 722~fb as calculated in Ref.~\cite{Frederix_2021}.

The production of $t\overline{t}Z/\gamma^{*}$ events was modeled using the \MGNLO generator.
The functional forms of  $\mu_\text{F}$ and $\mu_\text{R}$ were set to
\(0.5 \times \sum_i \sqrt{m^2_i+p^2_{\text{T},i}}\),
where the sum runs over all the particles generated in the ME calculation.
The cross-sections were calculated at NLO QCD and NLO EW accuracy using
\MGNLO as reported in Ref.~\cite{deFlorian:2016spz}.
The cross-section of \ttZ$(\ell\ell)$ with an off-shell $Z$ boson, denoted by \ttll, was scaled by an off-shell correction estimated at one-loop level in \alphas.
Two alternative $t\overline{t}Z/\gamma^{*}$ samples were simulated at NLO QCD with the \MGNLO generator and interfaced either to \PYTHIA[8] or \HERWIG[7.2].

The production of \ttH events was modeled using the
\POWHEGBOX[v2]~\cite{powhegtt,Nason:2004rx,Frixione:2007vw,Alioli:2010xd,Hartanto:2015uka}
generator. The functional form of  $\mu_\text{F}$ and $\mu_\text{R}$ was
set to \(\sqrt[3]{m_\text{T}(t)\cdot m_\text{T}(\bar{t}) \cdot m_\text{T}(H)}\).
The cross-section was calculated at NLO accuracy as reported in Ref.~\cite{deFlorian:2016spz}.
Two alternative \ttH samples were simulated, one with the \MGNLO generator interfaced to \PYTHIA[8] and one with the \POWHEGBOX generator interfaced to \HERWIG[7.2].

Diboson final states (\(VV\)) were simulated with the
\SHERPA[2.2.2] generator using the \NNPDF[3.0nnlo]~\cite{Ball:2014uwa} PDF set, taking off-shell effects into account.
MEs at NLO (LO) accuracy in QCD were used for up to one (three) additional parton emission(s).
Loop-induced processes were simulated at LO for up to one additional parton emission.
The calculations were matched and merged with the \SHERPA parton shower based on Catani--Seymour
dipole factorization using the MEPS@NLO prescription.
The virtual QCD corrections were provided by the \OPENLOOPS library.

The \ttbar production was simulated with \POWHEGBOX[v2] and normalized to the cross-section prediction at next-to-next-to-leading order (NNLO)
in QCD including the resummation of next-to-next-to-leading logarithmic (NNLL) soft-gluon terms calculated using
\TOPpp[2.0]~\cite{Beneke:2011mq,Cacciari:2011hy,Baernreuther:2012ws,Czakon:2012zr,Czakon:2012pz,Czakon:2013goa,Czakon:2011xx}.
The $h_{\textrm{damp}}$ parameter that controls the transverse momentum of the first gluon emission beyond the Born configuration in \POWHEGBOX was set to 1.5~$m_{\textrm{top}}$~\cite{ATL-PHYS-PUB-2016-020}.

The associated production of single top quarks with \(W\) bosons (\(tW\)) was modeled with the
\POWHEGBOX[v2] generator at NLO in QCD\@.
The diagram-removal scheme~\cite{Frixione:2008yi} was used to remove interference and overlap with \ttbar production.
The inclusive cross-section was corrected to the theory prediction calculated at NLO in QCD with NNLL soft-gluon corrections~\cite{Kidonakis:2010ux}.
Single top-quark production in the $s$-channel and $t$-channel was modeled with the \POWHEGBOX[v2] generator at NLO in QCD using the five- and four-flavor scheme, respectively.
The inclusive cross-section for the single-top $t$-channel process was corrected to the theory prediction calculated at NLO in QCD with
\HATHOR[2.1]~\cite{Aliev:2010zk,Kant:2014oha}.

The contribution from material photon conversions (Mat Conv) and internal photon conversion (Int Conv) were estimated by using $Z \to \ell^+\ell^-$ events, simulated with the \POWHEGBOX generator at NLO in QCD using the \CT[10nlo] PDF set interfaced to \PYTHIA[8] using the AZNLO tune~\cite{STDM-2012-23}.

The production of $t\bar t t\bar t$ events was simulated with the \MGNLO generator using the \NNPDF[3.1nlo] PDF set and was interfaced to \PYTHIA[8].
Several rare processes were also considered in the background estimate: $tZ$, $VH$, $t\bar{t}WW$, $t\bar{t}HH$, $t\bar{t}WH$, $t\bar{t}ZZ$, $tWZ$, $tWH, tHqb$, $ttt$, and $VVV$ production.
All processes involving at least one top quark were simulated with \MGNLO, while $VVV$ events were simulated with \SHERPA[2.2.2] and $VH$ was simulated with \POWHEGBOX[v2].

The simulated events were normalized to the cross-sections computed at the highest order available in perturbation theory.
Corrections were applied to the simulated events so that the selection efficiencies, energy scales
and energy resolutions of reconstructed objects match those determined from data control samples.


\section{Event reconstruction and object identification}
\label{sec:reco-sel}
Events of interest for this analysis are those compatible with the production of two top quarks with the same electric charge.
The dilepton final state studied here is characterized by the presence of two SS leptons, two
$b$-hadrons and missing transverse momentum from the two undetected neutrinos.

Interaction vertices are reconstructed from at least two tracks with transverse momentum ($\pt$) larger than $500~\mev$.
The primary vertex of the event is defined as the one with the largest $\sum_{\text{tracks}} \pt^2$ value~\cite{ATL-PHYS-PUB-2015-026}.
Events were selected using a combination of single-lepton and dilepton triggers.
Single-electron triggers required a minimum $\pt$ threshold of 24 (26)~$\gev$ in the 2015 (2016--2018)
data-taking period. Single-muon triggers had a lowest $\pt$ threshold of 20 (26)~$\gev$ in 2015 (2016--2018).
The dielectron triggers required two electrons with minimum $\pt$ thresholds ranging from 12~$\gev$ in 2015 to 24~$\gev$ in 2017--2018,
whereas the dimuon triggers used asymmetric $\pt$ thresholds for the leading (subleading) muons: 18 (8)~$\gev$ in 2015 and 22 (8)~$\gev$ in 2016--2018.
Finally, the electron+muon trigger required events to have an electron candidate with a 17~$\gev$ threshold and a
muon candidate with a 14~$\gev$ threshold for all periods.

Electron candidates are reconstructed from energy clusters in the electromagnetic calorimeter
matched to a track in the ID~\cite{EGAM-2018-01}.
They are required to satisfy $\pt>10~\GeV$ and $|\eta_{\mathrm{cluster}}|<2.47$, excluding the
pseudorapidity region corresponding to the transition between the barrel and endcap calorimeters
($1.37<|\eta_{\mathrm{cluster}}|<1.52$).
Loose and tight electron identification working points (WPs) are used~\cite{EGAM-2018-01}.

Muon candidates are reconstructed by combining tracks in the ID with tracks in the MS~\cite{MUON-2018-03}.
They are required to satisfy $\pt > 10~\gev$ and $|\eta| < 2.5$.
Loose and medium muon identification WPs are used~\cite{MUON-2018-03}.

Only lepton candidates with ID tracks originating from the primary vertex of the interaction are considered.
Electron (muon) candidates are required to have a transverse impact parameter, $d_0$, satisfying $|d_0/\sigma(d_0)|<5\,(3)$,
where $\sigma(d_0)$ is the measured uncertainty in $d_0$,
and a longitudinal impact parameter, $z_0$, satisfying $|z_0 \sin\theta|<0.5$~mm ($0.5$~mm),
where $\theta$ is the polar angle of the track.

To reject objects mistakenly identified as prompt leptons, namely misidentified jets and leptons from heavy-flavor hadron decays or photon conversions,
lepton candidates are also required to fulfill track and calorimeter-based isolation criteria~\cite{EGAM-2019-01,MUON-2018-03}.
Additionally, a boosted decision tree (BDT) discriminant, developed for Ref.~\cite{HIGG-2017-02}, referred to as the non-prompt lepton BDT, is employed to reject non-prompt leptons.
These leptons mainly originate from the decay of $c$- and $b$-hadrons
that can be identified by their track isolation and origin in displaced vertices, due to the decay of long-lifetime hadrons.
Two working points, named \textit{Tight} and \textit{VeryTight}, are defined based on the BDT output score.
The \textit{Tight} WP selects prompt leptons with an efficiency for muons (barrel/endcap electrons) that satisfy
the calorimeter- and track-based isolation criteria of about 60\% (60/70\%) for $\pt \sim 20~\gev$ and reaches
a plateau of 95\% (95/90\%) at $\pt \sim 40\ (40/65) ~\gev$.
The prompt lepton efficiency of the \textit{VeryTight} WP for muons (barrel/endcap electrons) that satisfy the calorimeter-
and track-based isolation criteria is about 55\% (55/60\%) for $\pt \sim 20~\gev$ and reaches
a plateau of 90\% (85/83\%) at $\pt \sim 40\ (40/65) ~\gev$.
The corresponding rejection factor against muons (electrons) from the decay of $b$-hadrons
ranges from 33 to 50 (20 to 50) for the \textit{Tight} WP, and from 50 to 100 (33 to 66) for the \textit{VeryTight} WP,
depending on \pt and $|\eta|$, after resolving ambiguities between overlapping reconstructed objects.

To reject background electrons due to a wrong charge assignment, a BDT based on calorimeter and track reconstruction parameters
is used, obtaining an efficiency of approximately 96\% in the barrel region and 81\% in the endcaps,
with corresponding rejection factors of 19 and 40 in the two regions, respectively.
Electrons originating in the detector material or from photon conversions are suppressed by rejecting candidates that are
associated with a displaced vertex at radial distance from the interaction point (IP) $r > 20$~mm and an
invariant mass of their combination with their closest opposite-charge tracks smaller than 100~\MeV.
The definitions of the lepton categories used in the analysis to define signal or control regions are summarized in Table \ref{tbl:defleps}
and comprise loose inclusive ($L_\text{inc}$), medium inclusive ($M_\text{inc}$), medium (\textit{M}), and tight (\textit{T}) leptons.\footnote{The \textit{M} and \textit{T} WPs are mutually exclusive while the $M_\text{inc}$ WP is the disjoint union of \textit{M} and \textit{T}.}

\begin{table}
\begin{center}
\caption{Summary of the definitions of the loose inclusive ($L_\text{inc}$), medium inclusive ($M_\text{inc}$), medium (\textit{M}), and tight (\textit{T}) leptons.}
\label{tbl:defleps}
\resizebox{\textwidth}{!}{
\begin{tabular}{l|c|c|c|c|c|c|c|c}
\toprule
Lepton & \multicolumn{4}{c|}{$e$} & \multicolumn{4}{c}{$\mu$} \\
\midrule
Category  & $L_\text{inc}$ & $M_\text{inc}$ & \textit{M} & \textit{T} &  $L_\text{inc}$ & $M_\text{inc}$ & \textit{M} & \textit{T} \\
\hline
Isolation                          &  \multicolumn{4}{c|}{Yes} & \multicolumn{4}{c}{Yes} \\
\hline
Non-prompt lepton BDT WP           &  No & \textit{Tight} & \textit{Tight-not-}  & \textit{VeryTight} &  No & \textit{Tight} & \textit{Tight-not-}  & \textit{VeryTight}  \\
&  &   & \textit{VeryTight} & &  &  & \textit{VeryTight} &  \\
\hline
Identification                     & Loose & \multicolumn{3}{c|}{Tight} & \multicolumn{4}{c}{Medium} \\
\hline
Electron charge-misassignment veto &  No & \multicolumn{3}{c|}{Yes} & \multicolumn{4}{c}{Not applicable} \\
Electron conversion candidate veto &  No & \multicolumn{3}{c|}{Yes} & \multicolumn{4}{c}{Not applicable} \\
\hline
$|d_0|/\sigma_{d_0}$               &  \multicolumn{4}{c|}{$<5$} & \multicolumn{4}{c}{$<3$ } \\
\hline
$|z_0 \sin \theta|$                &  \multicolumn{8}{c}{$< 0.5$ mm} \\
\bottomrule
\end{tabular}
}
\end{center}
\end{table}

Jets are reconstructed from collections of tracks in the ID and energy deposit clusters in the calorimeters,
combined using a particle-flow (PFlow) algorithm~\cite{PERF-2015-09}.
Jet candidates built from PFlow objects using the anti-$k_t$ algorithm~\cite{Cacciari:2008gp,Cacciari:2011ma}
with a radius parameter of $R=0.4$ are calibrated using simulation with corrections obtained from in situ techniques in
data~\cite{JETM-2018-05}. Jets are required to satisfy $\pT> 25$~\GeV and $|\eta|<2.5$.
A jet-vertex tagger (JVT)~\cite{JETM-2020-03} algorithm is applied to jets with $\pT< 60$~\GeV and $|\eta|<2.4$
to reduce pileup.
Jets containing $b$-hadrons are identified ($b$-tagged) via the DL1r algorithm~\cite{FTAG-2019-07}
that makes use of the impact parameters of tracks and the displaced vertices reconstructed in the ID, combined with
discriminating variables constructed with a recurrent neural network (NN)~\cite{FTAG-2019-07},
which exploits the spatial and kinematic correlations between tracks originating from the same $b$-hadron.
Three $b$-tagging WPs, corresponding to average $b$-tagging efficiencies of 60\%, 70\% and 77\%, are used.
The three WPs correspond to light-jet\footnote{Light jet refers to a jet not containing a $b$- or $c$-hadron.} rejection factors of about 2500, 700 and 200, and charm-jet ($c$-jet) rejection factors of about 40, 12 and 6,
as determined for jets with $\pt >20~\gev$ and $|\eta|<2.5$ in simulated $\ttbar$ events~\cite{FTAG-2019-07}.

To prevent leptons and jets being reconstructed multiple times as different objects in the event,
a sequential overlap removal procedure is applied to $L_\text{inc}$ leptons to remove such reconstruction ambiguities.
If two electrons are within $\Delta R=0.1$, only the one with the higher \pt\ is considered. If an electron and a muon are within $\Delta R=0.1$, the muon is removed if it is only reconstructed from an ID track and
the associated calorimeter deposit is consistent with a minimum ionizing particle, otherwise the electron is removed.
If an electron and a jet are within $\Delta R < 0.2$ of each other, the jet is removed if it is not $b$-tagged
(70\% WP) or if it has $\pt > 200~\gev$, otherwise the electron is removed.
Jets that are not $b$-tagged and contain less than three tracks with $\pt > 500~\mev$ and are within $\Delta R < 0.4$ from muons are rejected, otherwise, the muon is rejected.
For each lepton satisfying all previous criteria a variable-size cone of  $\Delta R < 0.04+10~\gev / p_{\mathrm{T,lep}}$
and up to a maximum $\Delta R=0.4$ is defined. If a selected jet that survived the previous steps is found within this cone, the lepton is rejected.

The missing transverse momentum, $\met$, is defined as the magnitude of the vector sum of the $\pt$ of all selected and
calibrated leptons and jets, plus a term to account for the momentum of soft particles that are not associated
with any of the selected objects~\cite{JETM-2020-03}.
This soft term is calculated from ID tracks matched to the selected primary vertex, enhancing its resilience to contamination from pileup interactions.

%


\section{Analysis workflow}
In this search for SS top-pair production only the decay topology where both $W$ bosons decay leptonically is considered.
Events are divided into non-overlapping phase-space regions. Four signal regions (SRs), nine control regions (CRs), and four validation regions (VRs), enriched in either signal or background events, are defined based on the predictions from the simulation of signal and background events in the ATLAS detector.

\begin{figure}[!htb]
\centering
\includegraphics[width=0.8\textwidth]{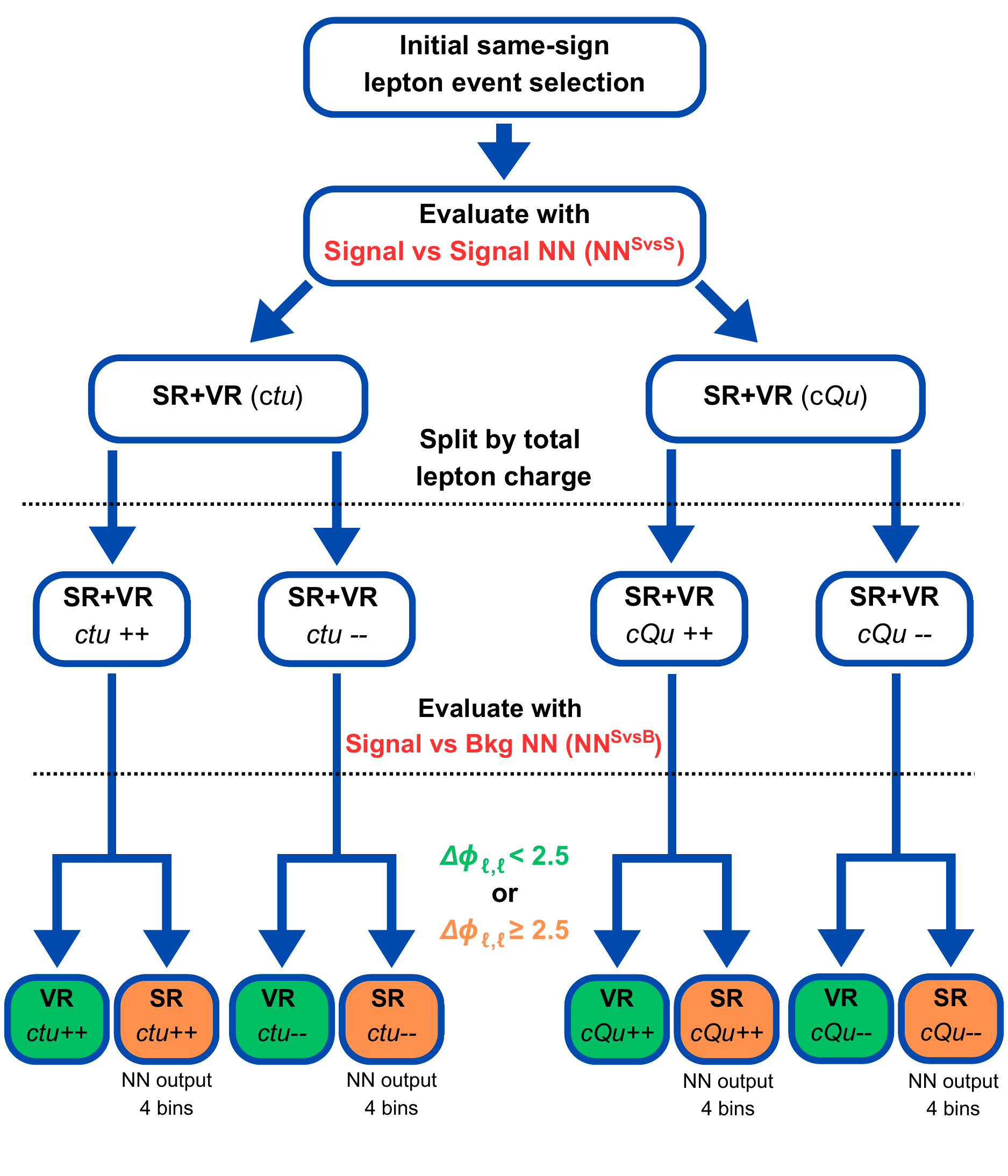}
\caption{Flowchart representation of the definition of signal regions (SRs) and validation regions (VRs). The first step is the selection of signal-like events. This is followed by the discrimination using a first set of deep neural networks, trained to separate signal events from the different EFT operators. The output of the NNs is used to define the SRs for $c_{tu}^{(1)}$ and $c_{Qu}$. Afterwards, the regions are split by total lepton charge into $++$ and $--$ regions. A second set of NNs is used to separate the signal from background contributions. As a last step, the regions are split based on the $\Delta \phi_{\ell,\ell}$ variable to define the final SRs and VRs. The output of this $\mathrm{NN^{SvsB}}$ is split into four bins per signal region.}
\label{fig:SR_sketch}
\end{figure}

Signal event candidates are pre-selected by requiring the presence of precisely two leptons, $\ell = e,\mu$, of the same electric charge and with $\pt > 20~\gev$. Events with more than two leptons are rejected.
At least two jets are required in the event. For the suppression of background contributions, either both leptons are classified as $M_\text{inc}$ and at least two jets satisfy the 77\% efficiency $b$-tagging WP, or both leptons are classified as \textit{T} and only one jet has to satisfy the 77\% WP\@. The dominant background processes in the SRs are the production of \ttW, \ttZ and \ttH. The \ttW and \ttZ normalizations are constrained by a likelihood fit to data described in Section~\ref{sec:results}. The \ttH is normalized as described in Section~\ref{sec:datamcsamples}.

The pre-selected events are classified using deep neural networks to create two separate regions enriched in events from the different EFT operators. A split by total lepton charge is applied to further divide the events into $++$ and $--$ regions, in order to be sensitive to the different rates and kinematic properties of $tt$ and $\bar{t}\bar{t}$ events
\footnote{Although the analysis is sensitive to both charge configurations, for the EFT interpretation in which the cross-section is highly asymmetric, the analysis gets most of its sensitivity from the $tt$ channel.}.
The SRs and VRs are defined by splitting the pre-selected events based on the azimuthal angle between the two charged leptons, $\Delta \phi_{\ell,\ell}$.
Figure~\ref{fig:SR_sketch} shows the definition of the SRs and VRs. In Figure~\ref{fig:New_VRs} the distribution of the sum of the \pt of the leptons ($H_{\mathrm{T}}^{\mathrm{lep}}$) and the distribution of the invariant mass of the two leptons ($m_{\ell\ell}$) is shown for the merged region (SRs+VRs) split by charge.
A second set of NNs, called $\mathrm{NN^{SvsB}}$, is used in the SRs to discriminate signal from background events.
The output of $\mathrm{NN^{SvsB}}$ is split into four bins per region, which are used in a maximum-likelihood fit to data.
The nine CRs are specifically designed to constrain different sources of background.
The fit is performed simultaneously in all SRs and CRs to extract the signal strength and several background normalization factors.
The VRs are used to validate the background modeling and are not used in the fit.

\begin{figure}[!htb]
\centering
\includegraphics[width=0.49\textwidth]{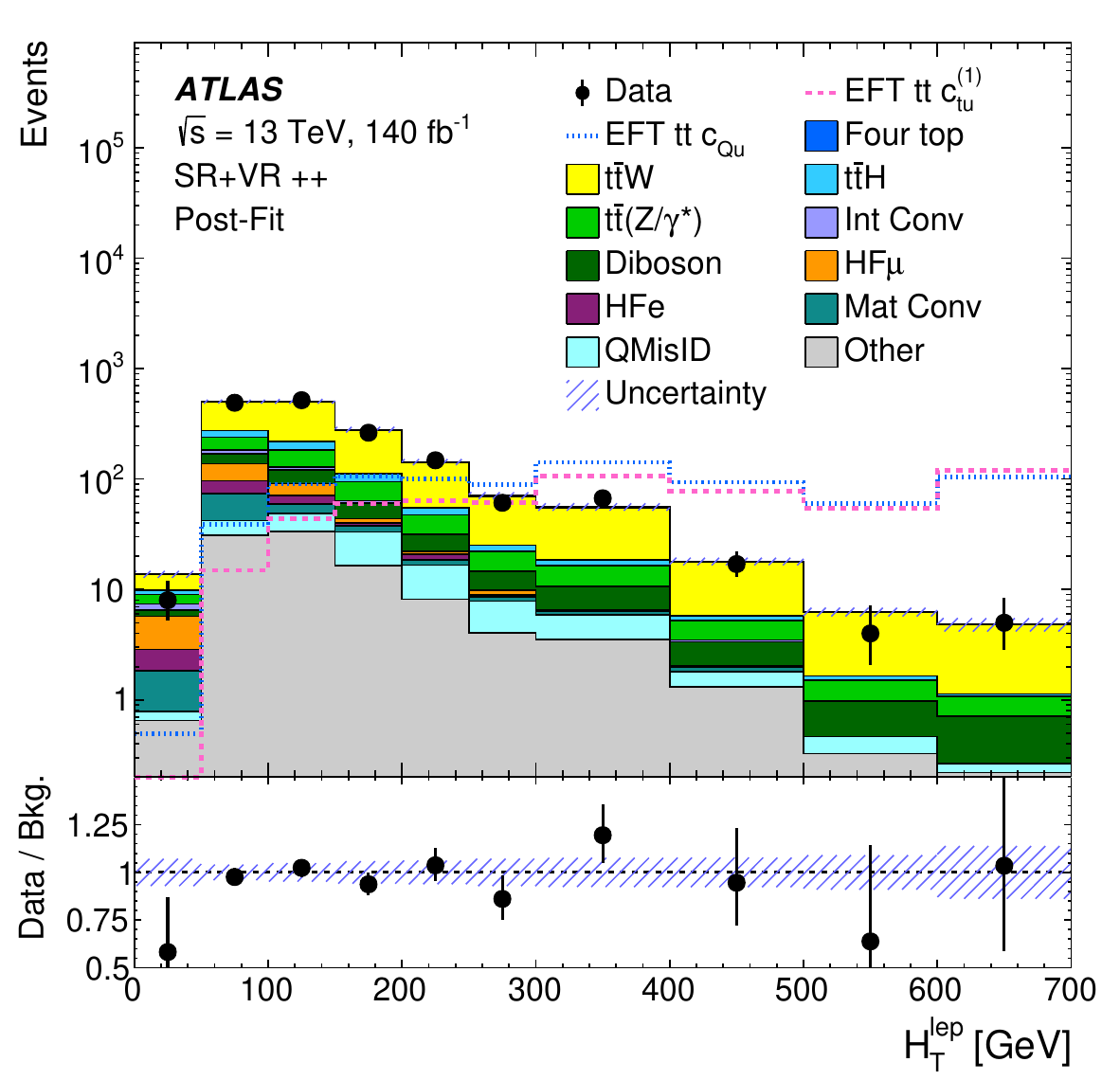}
\includegraphics[width=0.49\textwidth]{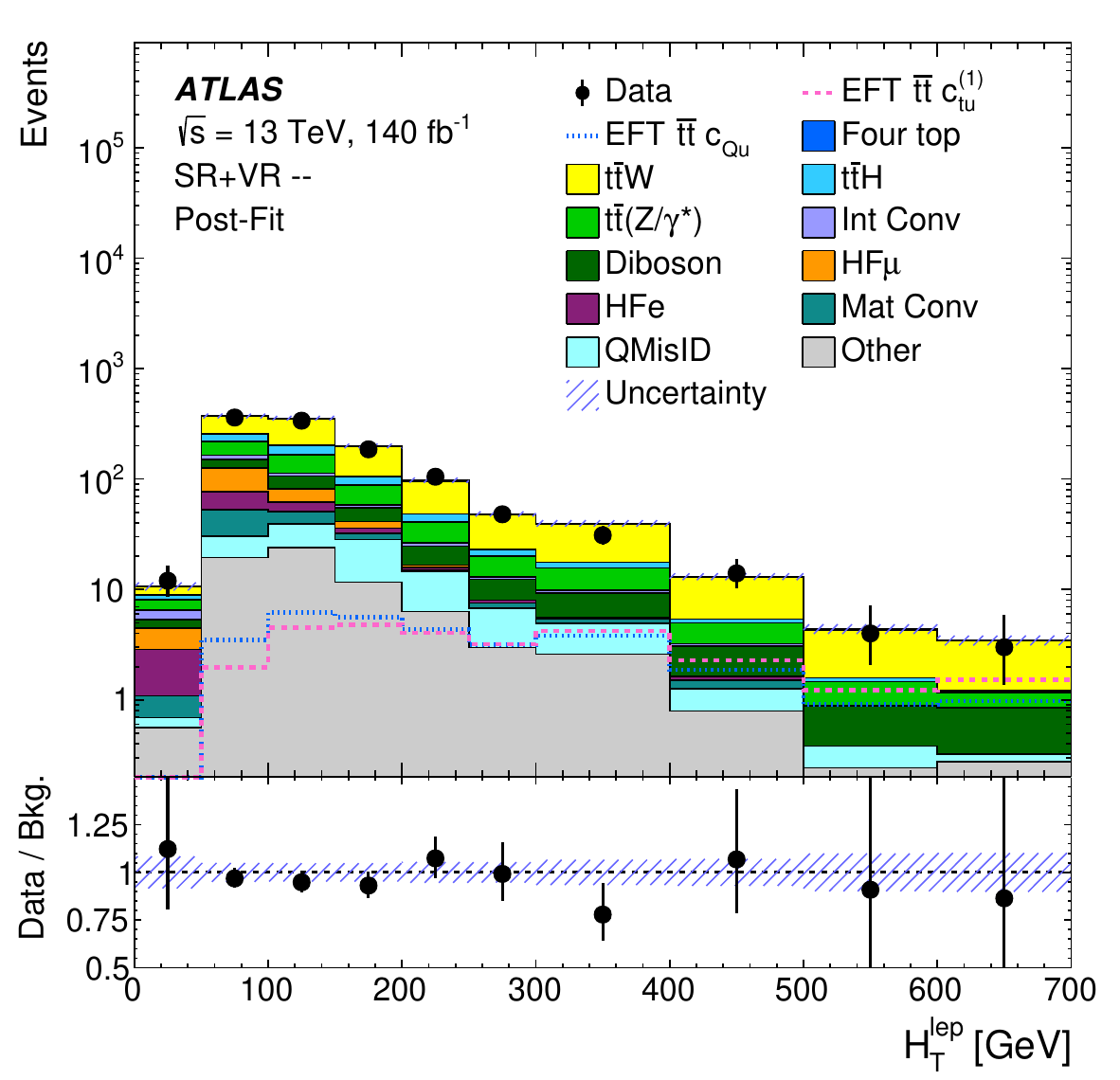} \\
\includegraphics[width=0.49\textwidth]{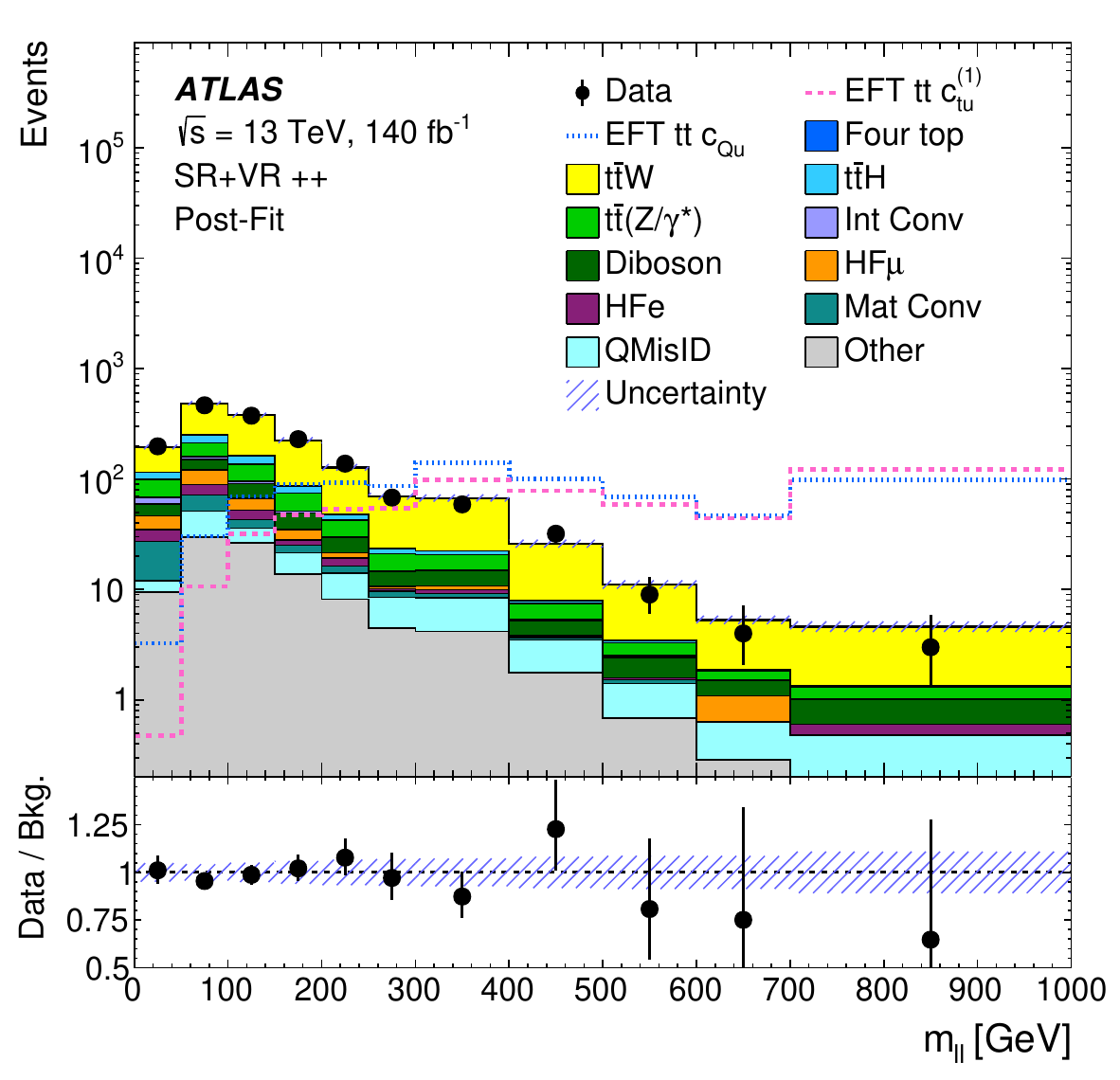}
\includegraphics[width=0.49\textwidth]{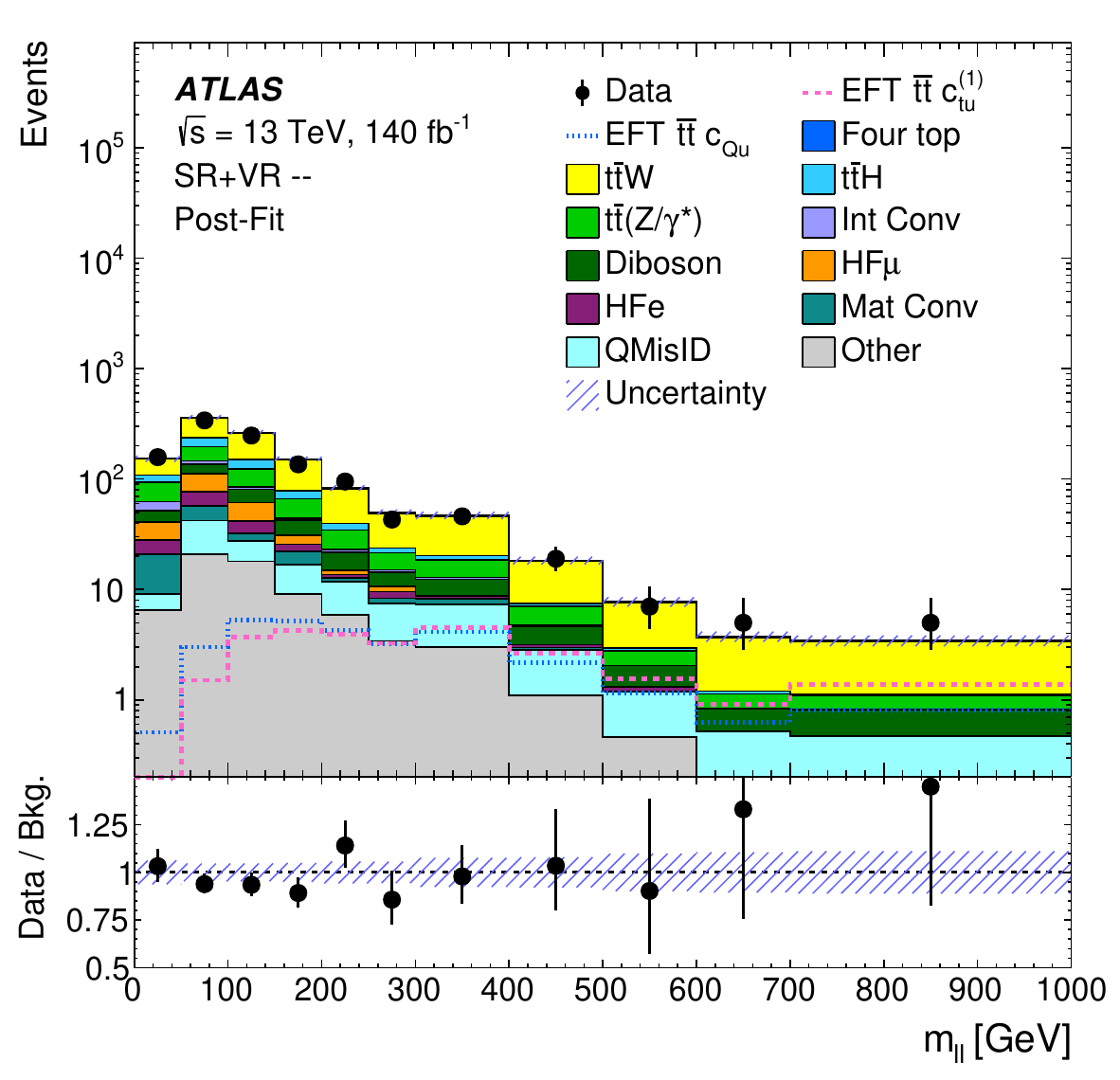} \\
\caption{Comparison between data and the background expectation of the $H_{\mathrm{T}}^{\mathrm{lep}}$ and $m_{\ell\ell}$ distributions in the merged regions (SRs+VRs) split by lepton charge, shown after the background-only likelihood fit. Two signal samples using the Wilson coefficient values of $c_{tu}^{(1)}=0.04$, $c_{Qu}^{(1)}=0.0$, $c_{Qu}^{(8)}=0.0$ and $c_{tu}^{(1)}=0.0$, $c_{Qu}^{(1)}=0.1$, $c_{Qu}^{(8)}=0.2$ are shown as a dotted line and a dashed line, respectively. They are normalized to their respective predicted cross-sections. The ratio of the data to the total post-fit background is shown in the lower panel. The combined statistical and systematic uncertainty in the MC simulation is indicated by the hatched band, while the vertical error bars represent the statistical uncertainty in the data. Overflow events are added to the last bin. \label{fig:New_VRs}}
\end{figure}

\subsection{Definition of signal enhanced regions via NNs}
\label{sec:DNN}

The two sets of NNs are trained based on the \textsc{Keras} library~\cite{chollet2015keras} with \textsc{Tensorflow}
as a backend~\cite{tensorflow2015-whitepaper} and the \textsc{Adam} optimiser~\cite{kingma2017adam}.
Both sets of NNs have the same architecture, with one input layer, five hidden layers consisting of 128, 64, 32, 16, and 8 nodes and a single node in the output layer.
The networks are trained with MC simulated samples. To make use of all the generated events,
the samples are split into two orthogonal subsets and a cross-training procedure is adopted
in which each subset is alternatively used for training and testing.

The operator ${\cal O}_{tu}^{(1)}$ and the two operators ${\cal O}_{Qu}^{(1)},~{\cal O}_{Qu}^{(8)}$
have different kinematic properties due the different chirality of the operators. The kinematic properties of the two ${\cal O}_{Qu}$ operators are very similar and therefore are combined into a single signal category.
A set of NNs, referred to as $\mathrm{NN^{SvsS}}$, are trained to discriminate between SS top-quark pairs generated by
the different EFT operators.
Signal events in the nominal \ensuremath{tt} and \ensuremath{\bar{t}\bar{t}} samples are weighted first according
to the EFT benchmark $c_{tu}^{(1)}=0.04,~c_{Qu}^{(1)}=0,~c_{Qu}^{(8)}=0$ and then according to
$c_{tu}^{(1)}=0,~c_{Qu}^{(1)}=0.1,~c_{Qu}^{(8)}=0.2$\footnote{It was found that using $c_{Qu}^{(8)}=0.2$ instead of $0.1$ for the NN training increases its discrimination power.}.
These two weighted samples are used to train and test the $\mathrm{NN^{SvsS}}$.
The classification is based on nine kinematic quantities:
the angular variables $\Delta \phi_{\ell,\ell}$, $\Delta R_{\ell,\ell}$, $\Delta \eta_{\ell,\ell}$ between the two leptons,
the invariant mass of the two-lepton system, the scalar sum of the \pt of all jets, the scalar sum of the \pt of all leptons,
the \pt of the leading jet, \met and the transverse mass of the combined leptons and \met system.
The $\mathrm{NN^{SvsS}}$ output distributions for the simulated signal and background events from MC simulations before the fit are shown in Figure~\ref{fig:NN_SvsS_Output}.
Lower or higher values correspond to ${\cal O}_{tu}^{(1)}$- or ${\cal O}_{Qu}$-like events, respectively.
Two orthogonal regions are defined by requiring $\mathrm{NN^{SvsS}}\leq0.538$ and
$\mathrm{NN^{SvsS}}>0.538$. This value corresponds to a classification efficiency of 65\% for both categories. The efficiency times acceptance values for signal events that enter the $SR_{tu}$ and $SR_{Qu}$ regions are 26.8\% and 12.4\% for signal events that originate from the ${\cal O}_{tu}^{(1)}$ operator, and 15.8\% and 19.5\% for signal events that originate from the ${\cal O}_{Qu}$ operators.

\begin{figure}[!htb]
\centering
\includegraphics[width=0.6\textwidth]{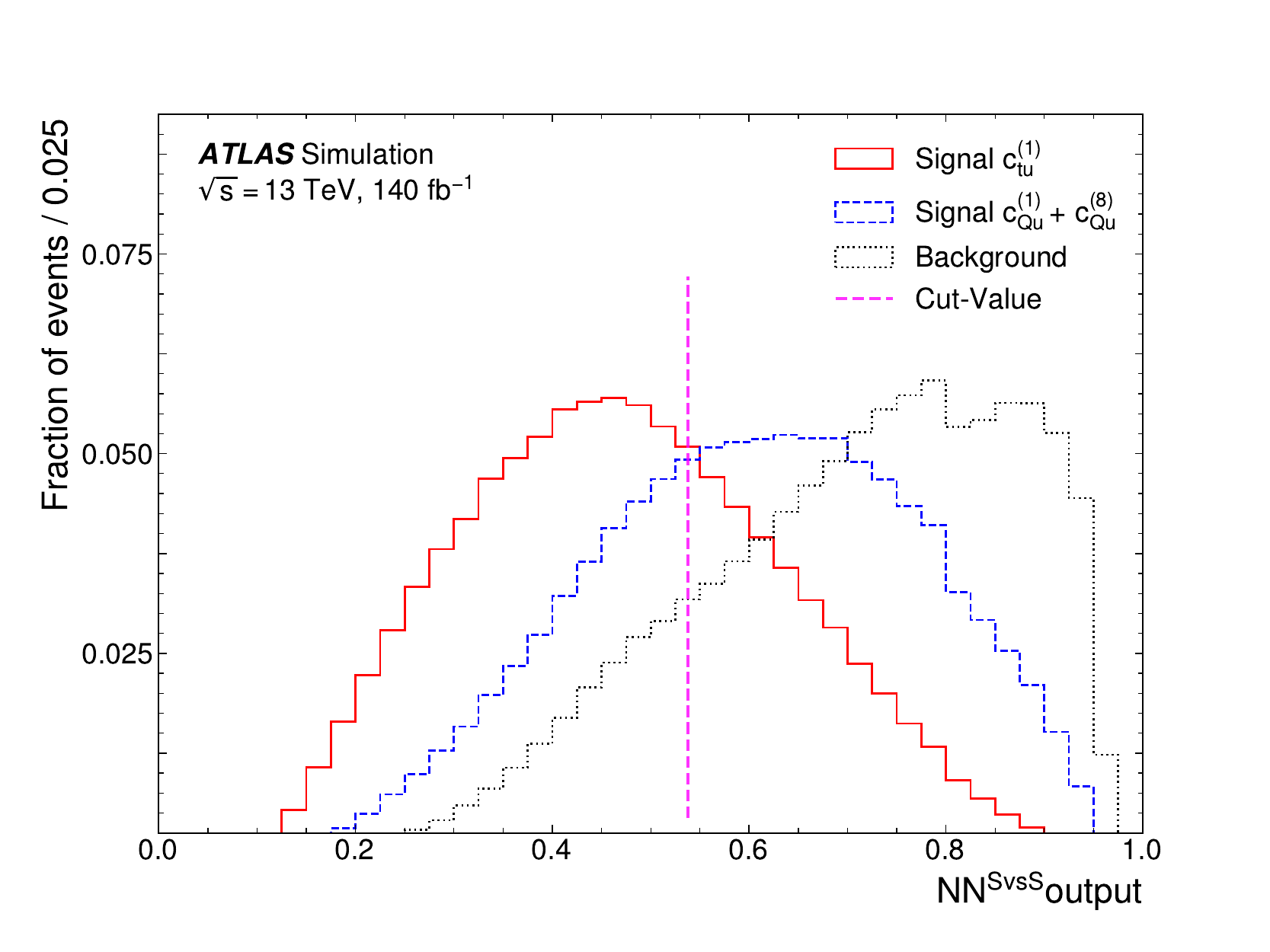} \\
\caption{NN output variable for $\mathrm{NN^{SvsS}}$ used to classify the data into $c_{tu}^{(1)}$- and $c_{Qu}$-like regions. The distribution for signal $c_{tu}^{(1)}$ was generated with $c_{tu}^{(1)}=0.04$ with the other Wilson coefficients being set to zero, while the other signal distribution was produced with $c_{tu}^{(1)}=0,c_{Qu}^{(1)}=0.1$ and $c_{Qu}^{(8)}=0.2$. The background distribution shown is the sum of all background processes from MC simulations before the likelihood fit.
The value chosen to separate the signal events is shown by the vertical dashed line. An efficiency of $\SI{65}{\percent}$ is obtained for the correct classification of $c_{Qu}$ events and of $c_{tu}^{(1)}$ events. \label{fig:NN_SvsS_Output}}
\end{figure}

A second set of NNs, referred to as $\mathrm{NN^{SvsB}}$, are used to discriminate the SS top-quark signal from background events.
The discrimination is based on six kinematic quantities:
the sum of the \pt of the leptons, the $b$-tagging score of the leading and sub-leading \pt jets,
the \pt of the leading jet, the transverse mass of the leptons and the \met system, and the number of jets in the event.
The distributions of these quantities show significant differences between the signal regions $SR_{Qu}$ and $SR_{tu}$
and for \ensuremath{tt} and \ensuremath{\bar{t}\bar{t}} events, therefore $\mathrm{NN^{SvsB}}$ is trained
separately for each of the four signal regions defined by the EFT operators and the charge of the two leptons:
$SR_{tu}^{++}$, $SR_{tu}^{--}$, $SR_{Qu}^{++}$ and $SR_{Qu}^{--}$.\footnote{Charge splitting of the signal samples is not applied in $\mathrm{NN^{SvsS}}$ training because it is found to be ineffective.}
Although $\Delta \phi_{\ell,\ell}$ and $\Delta R_{\ell,\ell}$ also show significant differences
between signal and background, these quantities are not used in the training of $\mathrm{NN^{SvsB}}$.
Instead, the value of $\Delta \phi_{\ell,\ell}$ is used after the $\mathrm{NN^{SvsB}}$ classification to define
the validation regions ($\Delta \phi_{\ell,\ell} < 2.5$) with a minor signal contamination
and to enhance the purity of the signal regions ($\Delta \phi_{\ell,\ell} \geq 2.5$).
Applying this cut results in the following rejection rates: 4.1\% (9.0 \%) of the signal events and 24.8\% (24.5\%) of the background events are rejected for $SR_{tu}^{++}$ ($SR_{tu}^{--}$), while 31.4\% (54.8\%) of the signal events and 85.7\% (85.9\%) of the background events are rejected for $SR_{Qu}^{++}$ ($SR_{Qu}^{--}$).

\subsection{Definition of background-enriched regions}
\label{sec:background}

The background contributions from different physics processes are modeled using MC simulations, data-driven techniques, or a combination of both.

The \ttW production is the dominant background in the SRs. Its contribution is modeled using MC events, as described in Section~\ref{sec:datamcsamples}. Dedicated analyses of \ttW production have measured larger cross-sections~\cite{TOPQ-2019-30} than predicted in theoretical calculations~\cite{Frederix_2021}. It is therefore crucial for this search to constrain the \ttW\ normalization in data. This is done using the $\mathrm{NN^{SvsB}}$ distributions in the SRs, which are enriched in \ttW events especially in the first bins where the SS top-quark pair contribution is negligible.

The normalization of the other backgrounds are either taken from simulation, highest order perturbation theory calculations or determined in dedicated CRs. In these CRs the different background contributions are enhanced through specific event selections that differ from those used to define the SRs.

Irreducible backgrounds are those which can produce two prompt SS leptons,
and include \ttV, $t\bar{t}H$, $tZ$, $VV$ and $t\bar{t}t\bar{t}$ production, with $V$ being a vector boson, $W$ or $Z$,
and rare processes. The rare processes are combined with the $tZ$ process in the category \textit{Other} in the following.
Their shape estimation relies on good modeling of these processes in the MC simulation.

The CRs defined for \ttZ and $VV$ share some common selection requirements. Events with three charged leptons in the final state, an opposite-sign (OS) pair with the same flavor plus an additional lepton of any flavor are selected. The invariant mass of the OS pair is required to be within 10~\GeV of the $Z$ boson mass.
The transverse momentum of the OS pair and of the SS pair are required to be $\pt (\text{OS}) > 10~\gev$ and $\pt (\text{SS}) > 20~\gev$, respectively.
Furthermore both SS leptons must be classified as $M_\text{inc}$ while the other lepton is required to be classified as $L_\text{inc}$.
At least four jets are required in the \ttZ CR, while events with exactly two or three jets are selected for the $VV$ CR\@.
Finally, in both CRs, at least one jet is required to be $b$-tagged according to the 60\% WP or at least two jets are required to be $b$-tagged according to the 77\% WP.

Reducible backgrounds contain at least one charge misidentified lepton or one fake or non-prompt lepton.
These backgrounds mainly come from \ttbar and $V$+jets production. Their shape is estimated from simulation, and the normalization is estimated by using data-driven procedures.
Their contributions depend on the lepton reconstruction, so they are estimated separately for electrons and muons.

Seven CRs enriched in non-prompt lepton events are defined.
Among these, five are used to estimate non-prompt leptons coming from the decay of $b$, $c$, and light hadrons,
referred to as heavy-flavor CRs, two for electrons (HF$e$) and three for muons (HF$\mu$).
In the heavy-flavor CRs, events with at least two jets and two SS leptons with $\pt > 20~\gev$ are selected.
Separate CRs are defined depending on the flavor of the sub-leading lepton.
Additionally, separate CRs are defined depending on the lepton classification according to the categories used in the analysis: namely the \textit{TM}, \textit{MT} and \textit{MM} combinations. This results in three CRs for each lepton flavor, namely HF$e$ \textit{TM}, HF$\mu$ \textit{TM}, HF$\mu$ \textit{MT}, HF$e$ \textit{MM} and HF$\mu$ \textit{MM}.
The CR for HF$e$ \textit{MT} is not used as some differences between the data and MC, which could be of statistical origin, are observed
in this region and as removing this CR from the fit has no impact on the expected result.
To suppress the contribution from \ttW production in these regions, only events with exactly one $b$-tagged jet (77\% WP) are selected and
in the \textit{TM} and \textit{MT} regions, the transverse mass of the lepton and \met system is required to be $m_{T,\ell\met}<250~\gev$.
The \pt distribution of the sub-leading lepton is used to extract the two normalization factors for $e$ and $\mu$,
by leaving them free in the fit to data.

Two CRs are defined to estimate the contributions to the background from leptons originating from photon conversions
$\gamma/\gamma^*\to e^+e^-$ in the detector material (CR Mat Conv) or in the vicinity of the interaction point (CR Int Conv).
Three-lepton events, where two leptons are of the same sign and have $\pt > 20~\gev$ and one lepton is of opposite sign with $\pt >10~\gev$ are selected. The flavor of the selected leptons is not directly restricted; instead, one of the same-sign leptons must be an electron to ensure the presence of a photon conversion candidate. Both SS leptons must be classified as $M_\text{inc}$ while the OS lepton must satisfy the $L_\text{inc}$ criteria.
Events with $b$-tagged jets (77\% WP) are rejected.
Events in which the invariant mass of the OS same-flavor leptons is within 10~\gev of the $Z$ boson mass are removed.
Furthermore, the invariant mass of the three-lepton system must satisfy $| m_{\ell\ell\ell} - m_Z | < 10~\gev$,
and $\met < 50~\gev$.
The two CRs require different selections on the invariant mass of the electron candidates and its closest track evaluated at the primary vertex, $m_{trk-trk,\text{PV}}$, and conversion vertices, $m_{trk-trk,\text{CV}}$. Specifically, material conversions candidates are required to have a conversion vertex found with a radial distance from the IP $r > 20~$mm and $0 < m_{trk-trk,\text{CV}} < 100~\mev$, while interaction conversion candidates have to be rejected as material conversion candidates and have $0 < m_{trk-trk,\text{PV}} < 100~\mev$.

An additional source of background is identified from electron charge misidentification (QMisID), mainly in \ttbar events.\footnote{The muon charge misidentification
rate is estimated to be of the order of $10^{-5}$ and has a negligible impact.}
This background is estimated with a data-driven method using $Z \to e e$ events and is negligible in the four $3\ell$ CRs. A likelihood-based method is used to measure the rate of reconstructed SS and OS electron pairs in specific control regions, while the non-$Z$ backgrounds are subtracted via a sideband method utilizing events outside the $Z$ boson mass window. The QMisID is then extracted from the ratio of SS to OS electron pairs in different \pt and $\eta$ bins and varies from $10^{-5}$ for low-$p_\text{T}$ electrons up to $4 \cdot 10^{-4}$ for high-$p_\text{T}$ electrons. The QMisID backgrounds are then estimated by applying these rates to data events that satify the requirements for the $2\ell$ regions, except that the two leptons are required to have opposite charges.


%
%
%

\section{Systematic uncertainties}
\label{sec:systematics}

The signal and background event yields in signal, validation and control regions are affected by uncertainties
due to the imperfect modeling of the physics processes and of the detector response. The systematic uncertainties are divided into two categories. The first category includes experimental uncertainties in the reconstruction of the final-state objects: leptons, jets, $b$-tagged jets and \met. The second category are modeling uncertainties, which are related to the choice of the MC generator, the PS and hadronization models, the choice of $\mu_\text{F}$ and $\mu_\text{R}$, and the uncertainties in the PDFs.

All systematic uncertainties are incorporated in the likelihood function used to extract the yields of
signal and backgrounds. Nuisance parameters (NPs) are associated with each source of systematic uncertainty
considered in the analysis. The nominal values and $\pm 1$ standard deviation ($\sigma$) variations of each NP are provided by dedicated studies.
The maximum-likelihood fit used to determine the signal and background yields also provides the values and uncertainties
of all NPs.

\subsection{Experimental uncertainties}
The uncertainty in the integrated luminosity is 0.83\%~\cite{DAPR-2021-01},
obtained using the LUCID-2 detector~\cite{LUCID2} for the primary luminosity measurements, complemented by measurements using the inner detector and the calorimeters.
The MC samples are reweighted to reproduce the pileup distribution observed in the data.
The corresponding pileup uncertainty is evaluated by varying these reweighting factors within their uncertainties, derived to cover the differences between the predicted and measured inelastic cross-section values~\cite{STDM-2015-05}.

Scale factors are applied to simulated events to account for differences between simulation and data in the efficiencies of electrons and muons when applying reconstruction, identification, trigger and isolation criteria.
These scale factors and their systematic uncertainties are evaluated for electrons and muons as a function of transverse energy (or \pt) and $\eta$ using tag-and-probe methods on $e^+ e^-$ and $\mu^+ \mu^-$ decays in $Z$ and $J/\psi$ events~\cite{MUON-2018-03,EGAM-2021-01,EGAM-2018-01}. The lepton energy/momentum scale and resolution and their corresponding uncertainties are also obtained from leptonic $Z$ decays~\cite{EGAM-2018-01,MUON-2022-01}.

The jet energy scale (JES) is calibrated using test-beam data, simulation, and in situ techniques~\cite{JETM-2018-05}. The corresponding uncertainty has 30 uncorrelated components, including effects due to pileup modeling, jet flavor composition and response, differences between jets from $b$-quarks and those from gluons or light quarks and effects of jets not fully contained in the calorimeter. The uncertainty in the jet energy resolution (JER) is evaluated via thirteen orthogonal components as a function of jet $p_T$ and $\eta$~\cite{JETM-2018-05}.

The uncertainty in the efficiency originating from the JVT requirement is obtained using $Z(\rightarrow \mu^+\mu^-)+$jets events~\cite{PERF-2014-03}. The uncertainty in \met due to possible miscalibration of its soft-track component is derived from data–simulation comparisons of the $p_T$ balance between the hard and soft $E_{\text{T}}^{\text{miss}}$ components~\cite{JETM-2020-03}.

The uncertainties related to the $b$-tagging efficiency of true $b$-jets and the mistagging rates of light-quark and $c$-jets is also considered. The $b$-tagging efficiency is measured in dileptonic $t\bar{t}$ events as a function of the jet \pt. Differences between data and simulation are corrected using scale factors. The uncertainty in these scale factors consists of 45 orthogonal components~\cite{FTAG-2018-01}. The rate of mistagging $c$-jets as $b$-jets is measured in semileptonic $t\bar{t}$ events, where one $W$ boson can decay to a $c$-quark and a down-type quark~\cite{FTAG-2020-08}. The mistagging rate of $c$-jets is calculated as a function of the jet \pt and has an overall uncertainty between 3\% and 17\%, broken down into 20 orthogonal components. The misidentification rate of light-quark jets is evaluated based on techniques described in Ref.~\cite{FTAG-2019-02}. The resulting uncertainties in the scale factors are also decomposed into 20 orthogonal components.

\subsection{Modeling uncertainties}
\label{sec:modeling}

For QCD \ttW, \ttZ, diboson, HF$e$, HF$\mu$, material photon conversions and internal photon conversions, the normalization is allowed to float freely in the fit. No normalization uncertainty is hence applied to $t\bar{t}$, single-top-quark, $Z+$jets and $W$+jets production, as all of these processes exclusively contribute to non-prompt lepton backgrounds. For all other processes, dedicated normalization uncertainties are included as NPs in the fit. For the additional contributions to \ttW from NLO electroweak diagrams a 20\% normalization uncertainty is applied~\cite{PhysRevD.103.094014}, as only QCD \ttW events are considered in the free normalization parameter. The \tttt, \ttH, and $tZ$ processes are assigned uncertainties of 20\%~\cite{Frederix:2017wme}, 11\%~\cite{deFlorian:2016spz}, and 20\%~\cite{TOPQ-2018-01}, respectively. An additional 50\% normalization uncertainty is assigned to EW \ttW, \ttZ, \ttH and \ttbar events containing additional heavy-flavor jets,
following Ref.~\cite{TOPQ-2018-05}. For the QmisID process a normalization uncertainty of 20\% is applied. All other minor processes are assigned a conservative 50\% normalization uncertainty.

For the dominant backgrounds, additional modeling uncertainties are considered, which use the alternative samples listed in parentheses in Table \ref{tab:mcconfig}. The comparison between different generator setups include acceptance effects, migration between event categories and effects on the kinematic distributions. The impact of the generator choice is used as a systematic uncertainty for \ttW and \ttH. For \ttW\ production this uncertainty is evaluated by comparing the nominal \Sherpa[2.2.10] sample with alternative samples generated with \MGNLO interfaced to \PYTHIA[8]. For \ttH the nominal \POWHEGBOX sample is compared with the alternative \MGNLO sample interfaced to \PYTHIA[8]. Uncertainties in the PS and hadronization modeling are assigned to the \ttW, \ttH and \ttZ processes. For \ttH and \ttZ, the uncertainties are derived by comparing the nominal samples with alternative samples for which the individual ME generators were interfaced to \HERWIG[7] instead of \PYTHIA[8]. For \ttW production, two alternative samples were generated. The matrix element of both samples was generated with \POWHEGBOX interfaced to \PYTHIA[8] for one sample and to \HERWIG[7] for the other. The relative differences between the two samples are determined for all bins in the analysis and applied to the nominal \SHERPA[2.2.10] \ttW\ sample.

Uncertainties related to the choice of $\mu_\text{F}$ and $\mu_\text{R}$ for the ME calculations are also taken into account for the
\ttH, \ttW, \ttZ, and $VV$ processes by varying their values by factors of 0.5 and 2.0 independently. For \ttW, separate scale uncertainties are derived for the QCD and EW components. For the signal samples, the effect of varying $\mu_\text{R}$ and $\mu_\text{F}$ is evaluated at generator level. Its effect on the jet multiplicity distribution is propagated via a reweighting procedure to the NN-output distributions in the SRs, resulting in a shape uncertainty of 2\%--25\% depending on the signal region and NN-output bin.

Uncertainties in initial-state radiation are taken into account by varying $\alpha_\text{s}^\text{ISR}$ according to the Var3c variation of the A14 tune~\cite{ATL-PHYS-PUB-2014-021} for \ttZ events. For \ttW, two additional PDF uncertainties are taken into account. The first uncertainty is defined by comparing the nominal PDF with $\alpha_\text{s}=0.118$ to variations with $\alpha_\text{s}=0.117$ and $\alpha_\text{s}=0.119$. The second uncertainty is based on the comparisons of the nominal PDF set \NNPDF[3.0nlo] to the alternative PDF sets \textsc{CT14nnlo}~\cite{Dulat:2015mca} and \textsc{MMHT2014nnlo68cl}~\cite{Harland-Lang:2014zoa}.

\subsection{Non-prompt-lepton background modeling uncertainties}

Non-prompt lepton background estimates are affected by uncertainties due to an imperfect modeling of the variables used in the
non-prompt lepton BDT\@. Systematic uncertainties, based on the residual disagreement observed between data and MC in the BDT input variables, are included as 34 NPs that can affect the shape of the distributions but not their normalization. Additional uncertainties, accounting for differences between the efficiencies of the non-prompt lepton BDT working points when using an alternative simulation, are considered. This uncertainty is estimated for the \textit{Tight-not-VeryTight} and \textit{VeryTight} WPs, respectively, by comparing the efficiencies in the nominal \POWPY[8] $t\bar{t}$ simulation with an alternative $t\bar{t}$ simulation (\POWHER[7.1.3] or \SHERPA[2.2.10]) as a function of the lepton $p_\text{T}$. As a consequence a normalization uncertainty of 20\% is applied to both HF$e$ and HF$\mu$ in the SRs.

Additional uncertainties are introduced to cover possible discrepancies between data and MC in the estimated rate of electron conversions.
These uncertainties are derived by comparing the conversion rates in data and MC in dedicated $2\ell$ validation regions, which require two \textit{Tight}
SS leptons where one of them is required to be a conversion candidate.
Interaction and material conversion extrapolation uncertainties of 50\% and 10\%, respectively, are assigned to the conversion rates in the SRs.

A systematic uncertainty is assigned to the QmisID background process. The uncertainty is assessed
by combining four different sources of uncertainty: the differences between misidentified electrons and positrons, the variation of rates within the $m_Z$ window,
the difference between the measured rates with the likelihood method and those obtained by matching reconstructed to generated   $Z \rightarrow e^+ e^-$ simulated events,
and the statistical uncertainty from the likelihood method. The uncertainty increases as a function of the electron $p_\text{T}$ and decreases with $|\eta|$ and ranges between 10\% and 60\%.


\section{Results}
\label{sec:results}

A maximum-likelihood fit is performed in the four SRs and the nine CRs to simultaneously determine the signal strength of the SS top-quark pair production and the background normalization factors defined in Section~\ref{sec:modeling}.
For the four SRs the $\mathrm{NN^{SvsB}}$ distributions are used as discriminants, while simpler discriminants are used in the CRs. The number of $b$-tagged jets is used for the \ttZ and diboson CRs, the transverse momentum of the sub-leading lepton is used for the HF$e$ TM and HF$\mu$ TM CRs, and for the remaining HF CRs and photon conversion CR the total number of events per region is used.

The likelihood function ${\cal L}(\mu,\vec{\lambda}, \vec{\theta})$, defined as the product of Poisson probability terms
for all bins, depends on the signal strength parameter $\mu$, the normalization factors for several backgrounds $\vec{\lambda}$,
and a set of NPs $\vec{\theta}$, each one linked to a source of systematic uncertainty described in Section~\ref{sec:systematics}.
Both $\mu$ and $\vec{\lambda}$ are allowed to vary freely in the likelihood fit.
The NPs $\vec{\theta}$ can produce variations in the signal and background expectations and are subject to Gaussian constraints in the likelihood fit.
Statistical uncertainties in each bin, due to the limited size of the experimental and simulated samples, are taken into account with dedicated parameters.
The statistical analysis is performed using the \textsc{RooFit} framework~\cite{RooFit} with the statistical model built following the HistFactory format~\cite{Cranmer:1456844}.

Figures~\ref{fig:CR_Summary_postfit} and \ref{fig:SR_Summary_postfit} show the observed and expected number of events in the nine CRs and four SRs, respectively, after the likelihood fit under the signal-plus-background hypothesis.
The corresponding post-fit yields are shown in Tables~\ref{tab:Asimov_CR_2L_Postfit}--\ref{tab:Asimov_SR_Postfit}.
They are consistent with the observed data within their uncertainties and align with the pre-fit yields. In the VRs good agreement between data and simulation is also observed post-fit.

\begin{figure}[H]
\centering
\includegraphics[width=0.85\textwidth]{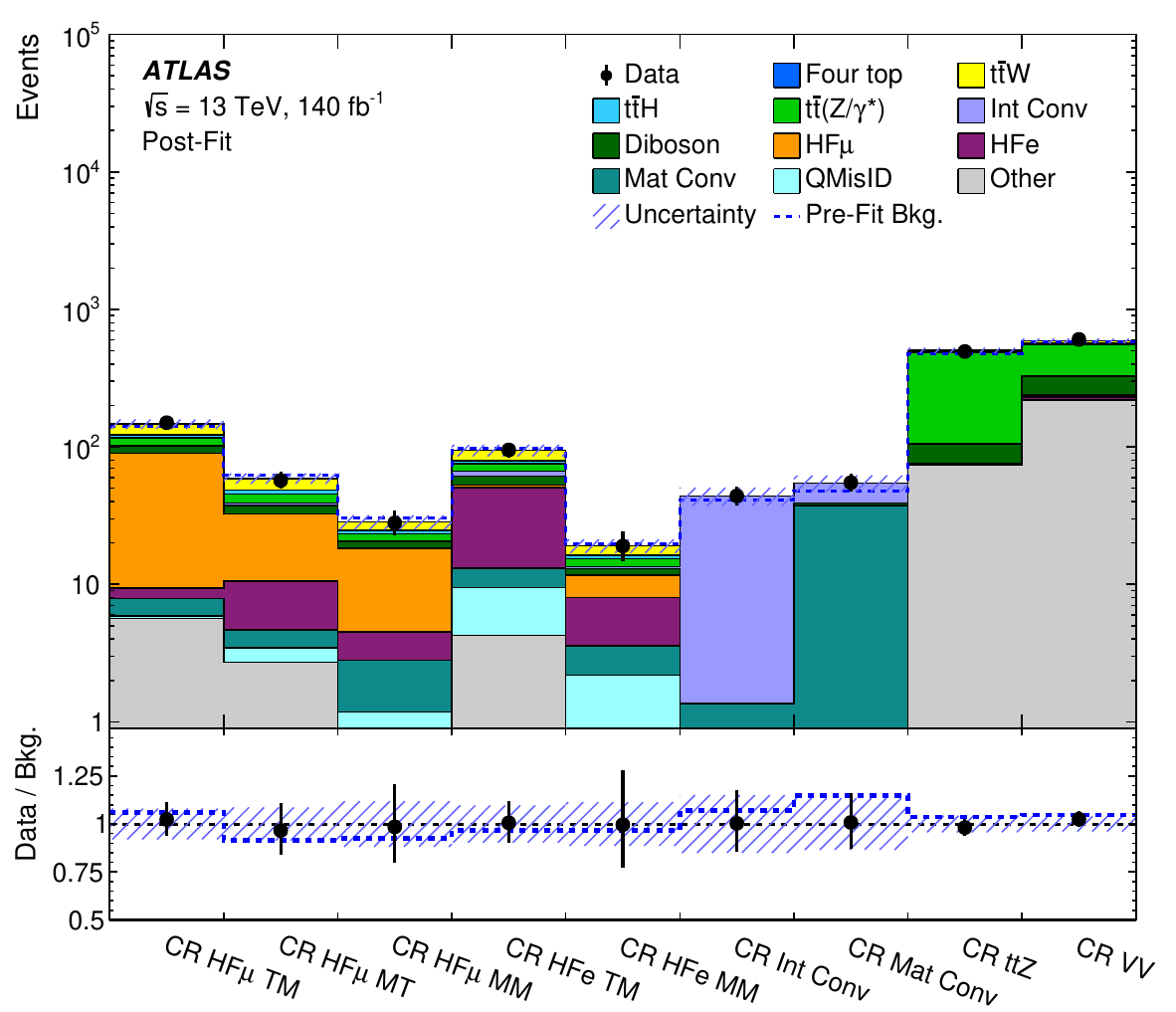} \\
\caption{Comparison between the event yields in data and the background expectation after the likelihood fit for the nine control regions. The post-fit background expectations are shown as filled histograms, the combined pre-fit background expectations are shown as dashed lines. The ratio of the data to the total post-fit background is shown in the lower panel. The combined statistical and systematic uncertainty in the simulation is indicated by the hatched band, while the vertical error bars represent the statistical uncertainty in the data. \label{fig:CR_Summary_postfit}}
\end{figure}

\begin{figure}[H]
\centering
\includegraphics[width=0.49\textwidth]{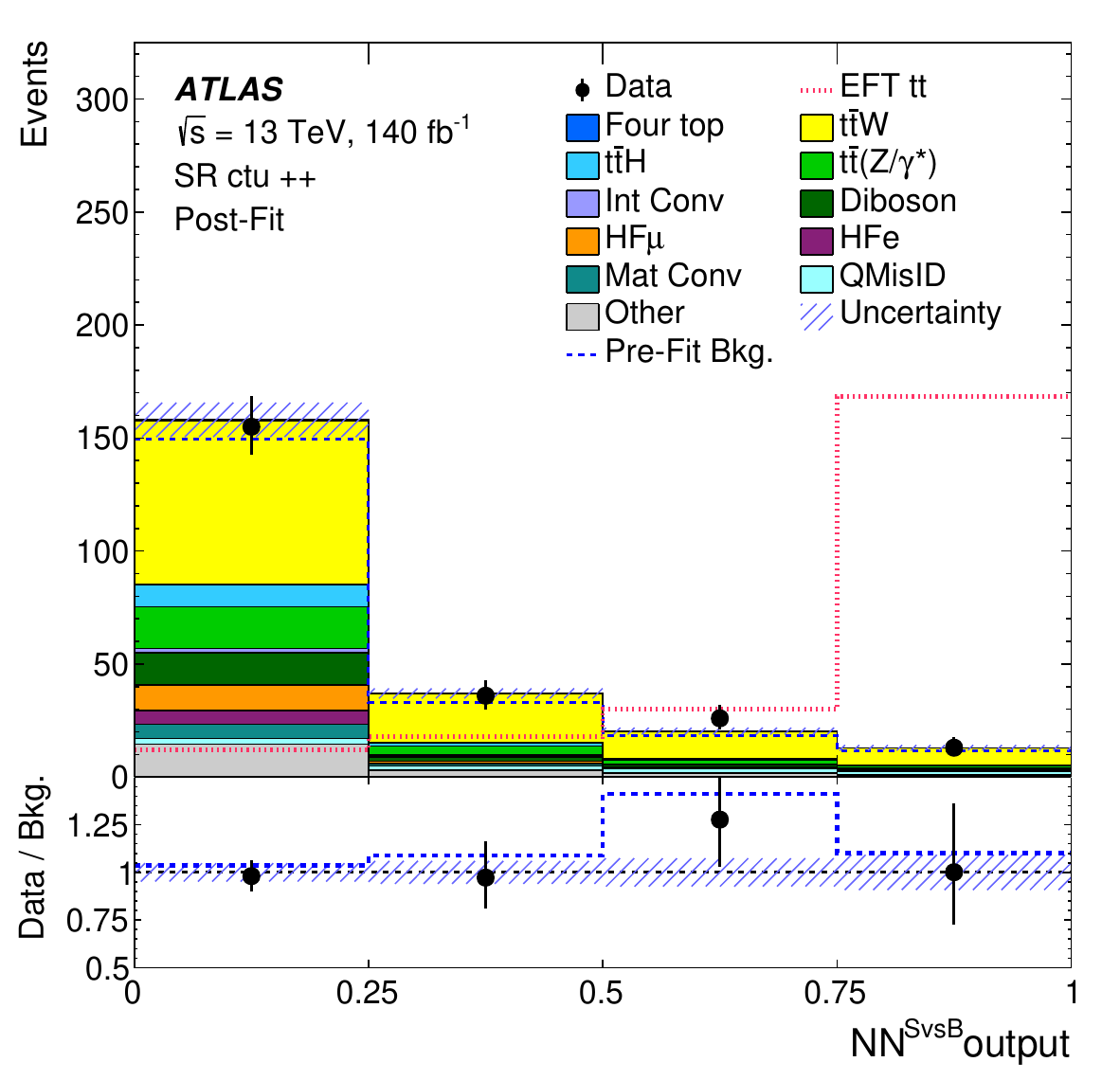}
\includegraphics[width=0.49\textwidth]{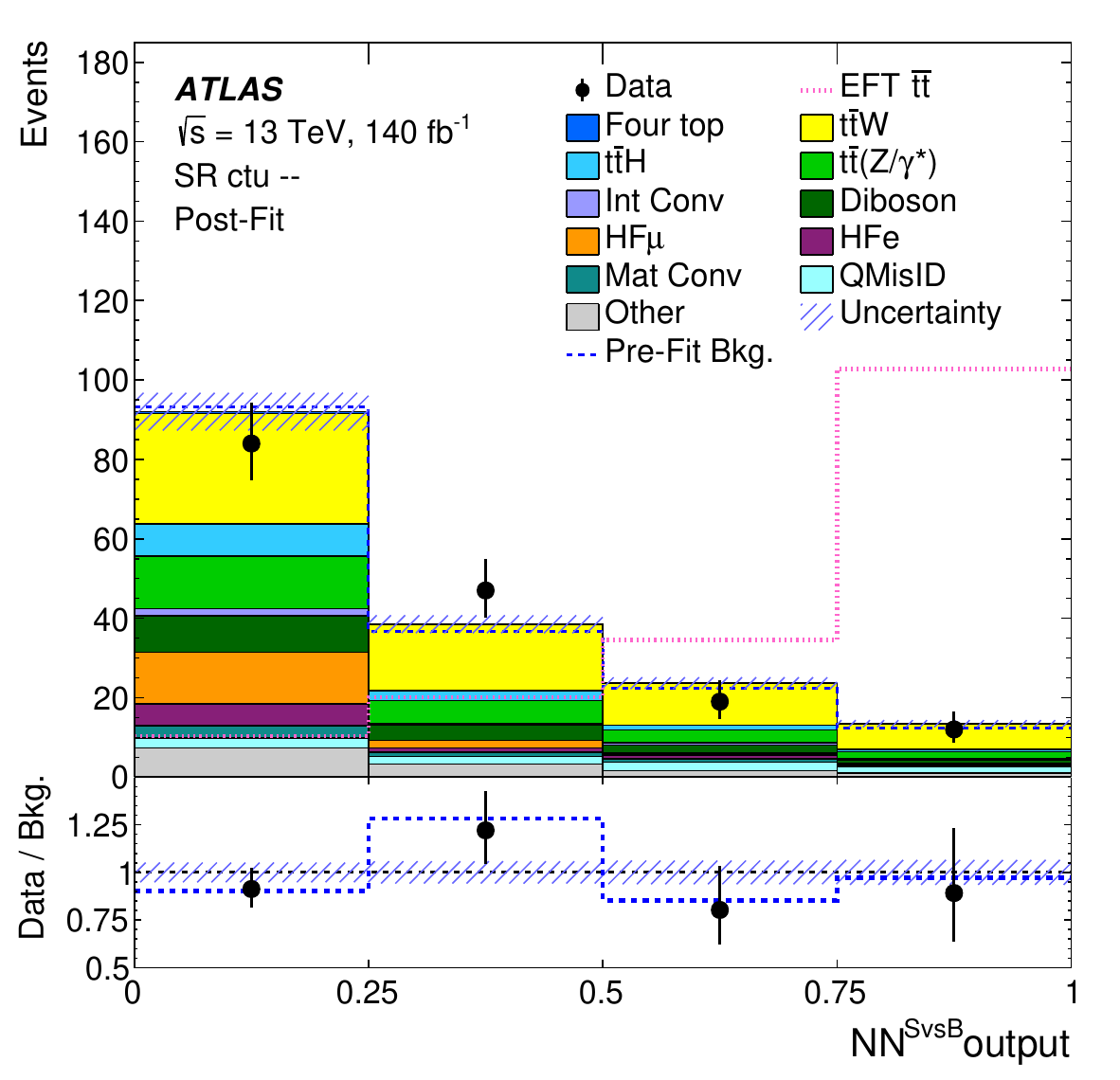} \\
\includegraphics[width=0.49\textwidth]{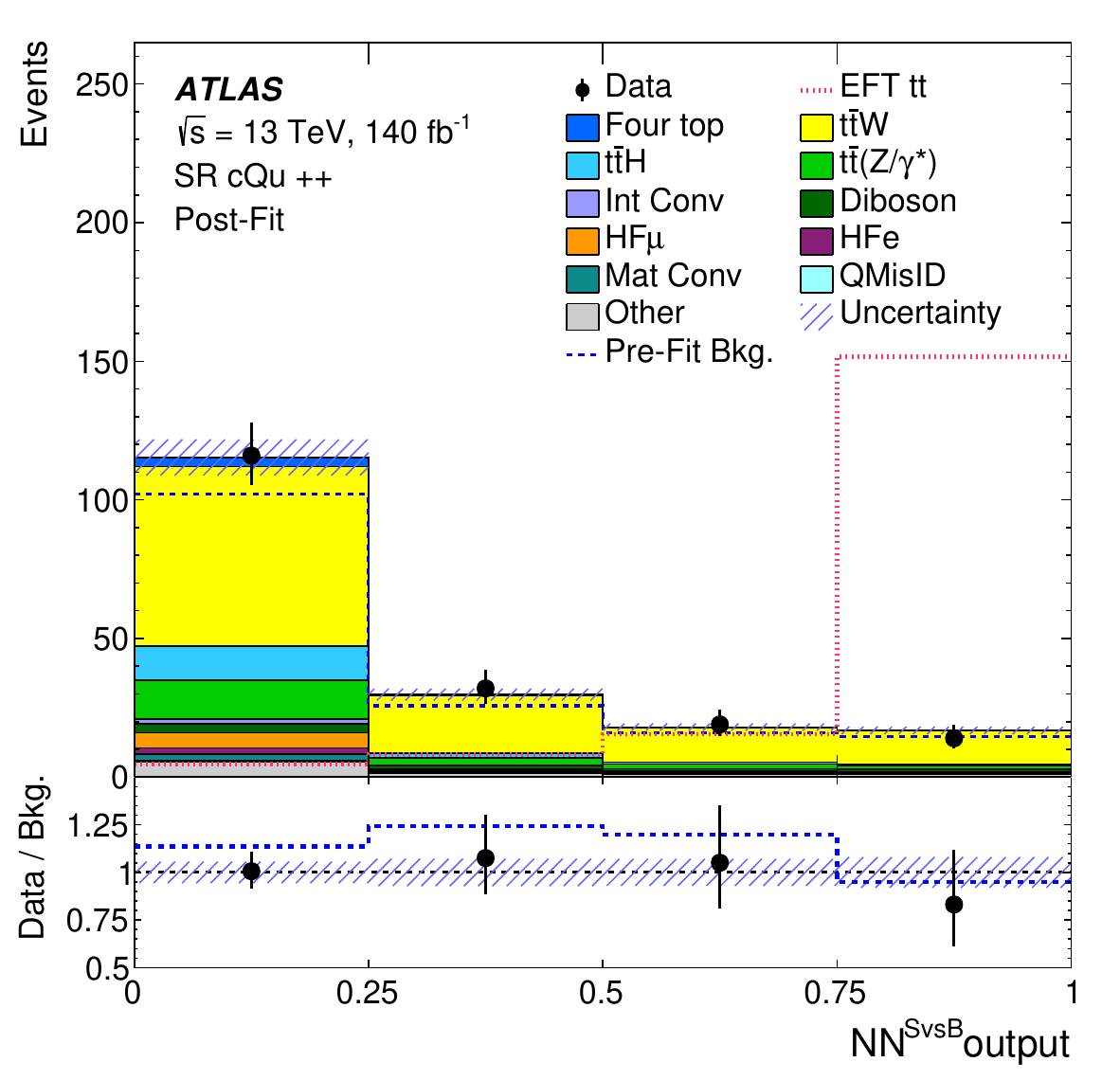}
\includegraphics[width=0.49\textwidth]{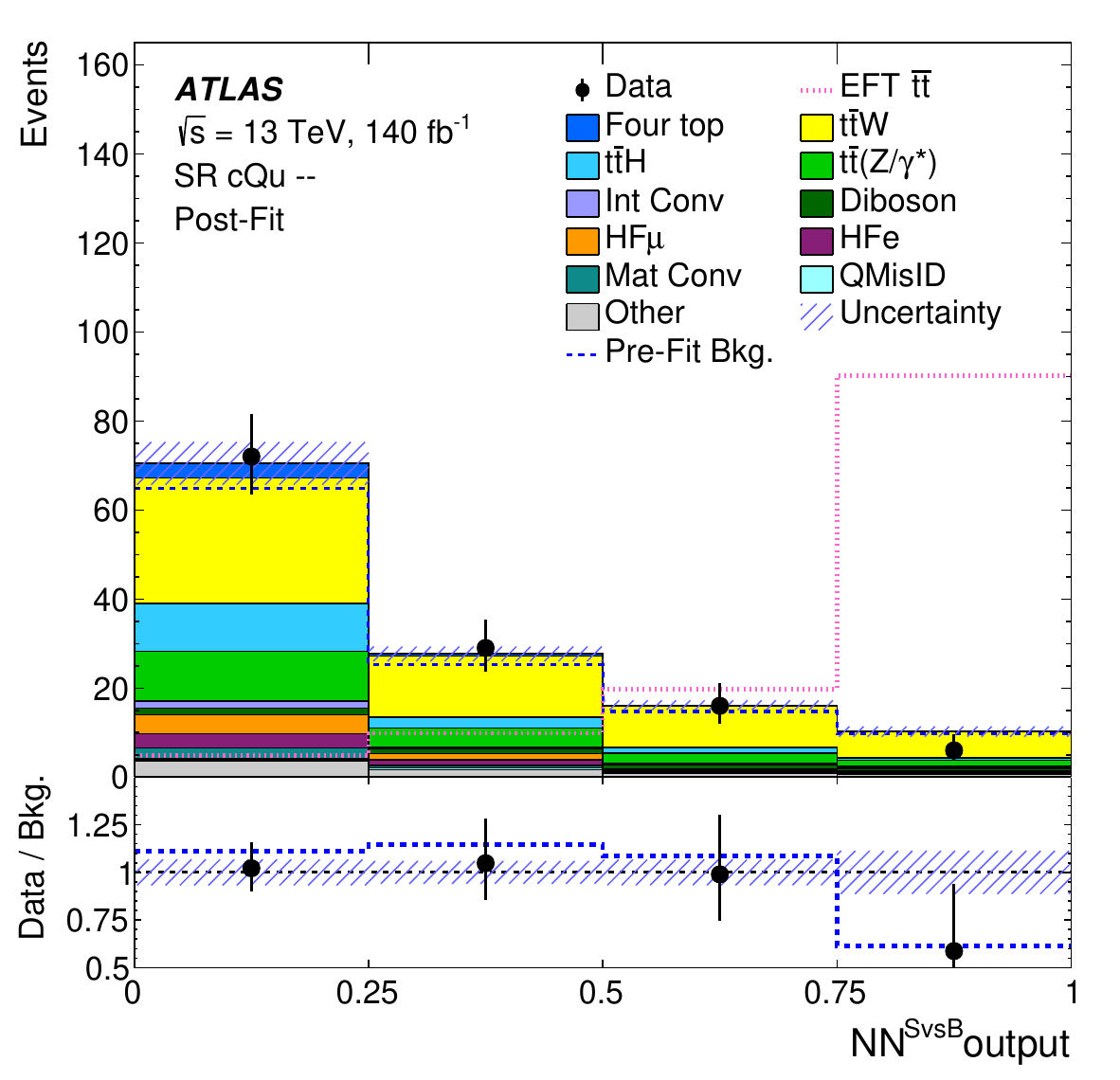} \\
\caption{Distributions of the $\mathrm{NN^{SvsB}}$ output for data and the expected background after the likelihood fit in the four signal regions. The post-fit background expectations are shown as filled histograms, the combined pre-fit background expectations are shown as dashed lines. The signal distribution using the Wilson coefficient values $c_{tu}^{(1)}=0.04$, $c_{Qu}^{(1)}=0.1$, $c_{Qu}^{(8)}=0.1$ is shown with a dotted line, normalized to the same number of events as the background. The ratio of the data to the total post-fit background is shown in the lower panel. The combined statistical and systematic uncertainty in MC simulation is indicated by the hatched band, while the vertical error bars represent the statistical uncertainty in the data. \label{fig:SR_Summary_postfit}}
\end{figure}

\begin{table}[H]
\centering
\caption{Summary of the observed and predicted number of events in the five 2$\ell$ control regions. The background prediction
is shown after the combined likelihood fit to data under the signal-plus-background hypothesis across all control regions and signal regions. The uncertainties in the total yields are smaller than the sum in quadrature of the uncertainties in the individual contributions due to anti-correlations resulting from the likelihood fit. \label{tab:Asimov_CR_2L_Postfit}}
\begin{tabular}{l
S[table-format=3.1]@{\hspace{0.45cm}$\pm$\hspace{-0.30cm}}S[table-format=4.1]
S[table-format=4.1]@{\hspace{0.45cm}$\pm$\hspace{-0.45cm}}S[table-format=4.1]
S[table-format=4.1]@{\hspace{0.45cm}$\pm$\hspace{-0.45cm}}S[table-format=4.1]
S[table-format=4.1]@{\hspace{0.45cm}$\pm$\hspace{-0.45cm}}S[table-format=4.1]
S[table-format=3.1]@{\hspace{0.45cm}$\pm$\hspace{-0.45cm}}S[table-format=4.1]
}
\toprule
Process & \multicolumn{2}{c}{$\text{CR HF}\mu \text{ TM}$} & \multicolumn{2}{c}{$\text{CR HF}\mu \text{ MT}$} & \multicolumn{2}{c}{$\text{CR HF}\mu \text{ MM}$} & \multicolumn{2}{c}{$\text{CR HF}e \text{ TM}$} & \multicolumn{2}{c}{$\text{CR HF}e \text{ MM}$} \\ \midrule
$t\bar{t}W$ & 24.0 &  4.9 & 10.3 &  2.0 & 3.73 &  0.86 & 15.1 &  2.9 & 2.76 &  0.59 \\
$t\bar{t}(Z/\gamma*)$ & 13.6 &  2.6 & 6.2 &  1.1 & 2.59 &  0.58 & 8.4 &  1.8 & 1.90 &  0.37 \\
$t\bar{t}H$ & 6.6 &  4.0 & 3.2 &  1.9 & 1.28 &  0.79 & 4.1 &  2.4 & 0.90 &  0.58 \\
$\text{Four top}$ & 0.113 &  0.028 & 0.071 &  0.017 & 0.046 &  0.012 & 0.069 &  0.019 & 0.036 &  0.009 \\
$\text{Diboson}$ & 11.9 &  6.1 & 4.9 &  2.5 & 2.2 &  1.1 & 8.6 &  4.4 & 1.35 &  0.71 \\
$\text{HFe}$ & 1.6 &  1.1 & 5.9 &  2.9 & 1.71 &  0.97 & 37 &  12 & 4.5 &  1.6 \\
$\text{$\text{HF}\mu$}$ & 80 &  14 & 21.9 &  5.6 & 13.7 &  3.2 & 2.20 &  0.66 & 3.62 &  0.99 \\
$\text{Mat Conv}$ & 2.0 &  7.1 & 1.20 &  0.56 & 1.62 &  0.51 & 3.7 &  2.1 & 1.38 &  0.43 \\
$\text{Int Conv}$ & 0.68 &  0.41 & 1.7 &  1.0 & 0.30 &  0.18 & 5.5 &  3.2 & 0.48 &  0.31 \\
$\text{QMisID}$ & 0.28 &  0.13 & 0.75 &  0.54 & 0.38 &  0.26 & 5.2 &  2.9 & 1.6 &  1.0 \\
$\text{Other}$ & 5.6 &  1.5 & 2.71 &  0.68 & 0.81 &  0.22 & 4.2 &  1.1 & 0.63 &  0.17 \\
\midrule
Total Bkg. & 147 &  12 & 59.0 &  5.1 & 28.4 &  3.4 & 94.4 &  9.2 & 19.1 &  2.2 \\
\midrule
Data & \multicolumn{2}{c}{150} & \multicolumn{2}{c}{57} & \multicolumn{2}{c}{28} & \multicolumn{2}{c}{95} & \multicolumn{2}{c}{19} \\
\bottomrule
\end{tabular}
\end{table}

\begin{table}[H]
\caption{Summary of the observed and predicted number of events in the four 3$\ell$ control regions. The background prediction
is shown after the combined likelihood fit to data under the signal-plus-background hypothesis across all control
regions and signal regions. The uncertainties in the total yields are smaller than the sum in quadrature of the
uncertainties in the individual contributions due to anti-correlations resulting from the likelihood fit. \label{tab:Asimov_CR_3L_Postfit}}
\centering
\begin{tabular}{l
S[table-format=4.1]@{\hspace{0.45cm}$\pm$\hspace{-0.45cm}}S[table-format=4.1]
S[table-format=4.1]@{\hspace{0.4cm}$\pm$\hspace{-0.45cm}}S[table-format=4.1]
S[table-format=4.1]@{\hspace{0.4cm}$\pm$\hspace{-0.35cm}}S[table-format=4.1]
S[table-format=4.1]@{\hspace{0.4cm}$\pm$\hspace{-0.335cm}}S[table-format=4.1]
}
\toprule
Process & \multicolumn{2}{c}{$\text{CR Int Conv}$} & \multicolumn{2}{c}{$\text{CR Mat Conv}$} & \multicolumn{2}{c}{$\text{CR ttZ}$} & \multicolumn{2}{c}{$\text{CR VV}$} \\ \midrule
$t\bar{t}W$ & \multicolumn{2}{c}{--} & \multicolumn{2}{c}{--} & 8.4 & 1.8 & 24.5 & 4.7 \\
$t\bar{t}(Z/\gamma*)$ & \multicolumn{2}{c}{--} & \multicolumn{2}{c}{--} & 378 & 32 & 230 & 28 \\
$t\bar{t}H$ & \multicolumn{2}{c}{--} & \multicolumn{2}{c}{--} & 10.0 & 6.3 & 6.3 & 4.0 \\
$\text{Four top}$ & \multicolumn{2}{c}{--} & \multicolumn{2}{c}{--} & 1.61 & 0.32 & 0.092 & 0.020 \\
$\text{Diboson}$ & 0.025 & 0.019 & 1.40 & 0.72 & 29 & 15 & 90 & 45 \\
$\text{HFe}$ & \multicolumn{2}{c}{--} & \multicolumn{2}{c}{--} & 0.47 & 0.35 & 9.2 & 6.8 \\
$\text{$\text{HF}\mu$}$ & \multicolumn{2}{c}{--} & \multicolumn{2}{c}{--} & 1.04 & 0.35 & 7.5 & 1.8 \\
$\text{Mat Conv}$ & 1.3 & 1.1 & 37.6 & 8.6 & 0.59 & 0.40 & 2.19 & 0.77 \\
$\text{Int Conv}$ & 42.5 & 6.8 & 15.6 & 4.3 & 0.14 & 0.15 & 1.66 & 0.96 \\
$\text{QMisID}$ & \multicolumn{2}{c}{--} & \multicolumn{2}{c}{--} & 0.22 & 0.17 & 0.83 & 0.41 \\
$\text{Other}$ & \multicolumn{2}{c}{--} & \multicolumn{2}{c}{--} & 74 & 23 & 218 & 41 \\
\midrule
Total Bkg. & 43.9 & 6.6 & 54.6 & 7.3 & 503 & 22 & 590 & 23 \\
\midrule
Data & \multicolumn{2}{c}{44} & \multicolumn{2}{c}{55} & \multicolumn{2}{c}{494} & \multicolumn{2}{c}{605} \\
\bottomrule
\end{tabular}
\end{table}

\begin{table}[H]
\caption{Summary of the observed and predicted number of events in the four signal regions. The background prediction
is shown after the combined likelihood fit to data under the signal-plus-background hypothesis across all control
regions and signal regions. The uncertainties in the total yields are smaller than the sum in quadrature of
the uncertainties in the individual contributions due to anti-correlations resulting from the likelihood fit. \label{tab:Asimov_SR_Postfit}}
\centering
\begin{tabular}{l
S[table-format=4.1]@{\hspace{0.4cm}$\pm$\hspace{-0.25cm}}S[table-format=4.1]
S[table-format=4.1]@{\hspace{0.4cm}$\pm$\hspace{-0.25cm}}S[table-format=4.1]
S[table-format=4.1]@{\hspace{0.4cm}$\pm$\hspace{-0.25cm}}S[table-format=4.1]
S[table-format=4.1]@{\hspace{0.4cm}$\pm$\hspace{-0.25cm}}S[table-format=4.1]
}
\toprule
Process & \multicolumn{2}{c}{$SR_{ctu ++}$} & \multicolumn{2}{c}{$SR_{ctu --}$} & \multicolumn{2}{c}{$SR_{cQu ++}$} & \multicolumn{2}{c}{$SR_{cQu --}$} \\ \midrule
$t\bar{t}W$ & 114 & 15 & 62 & 10 & 110 & 15 & 56.9 & 8.9 \\
$t\bar{t}(Z/\gamma*)$ & 25.5 & 3.0 & 24.1 & 2.6 & 19.5 & 2.0 & 19.1 & 2.1 \\
$t\bar{t}H$ & 12.4 & 7.5 & 12.3 & 7.1 & 15.1 & 9.6 & 15.1 & 9.2 \\
$\text{Four top}$ & 0.72 & 0.15 & 0.69 & 0.14 & 4.16 & 0.83 & 4.07 & 0.82 \\
$\text{Diboson}$ & 18.1 & 9.3 & 15.9 & 8.1 & 6.3 & 3.2 & 4.2 & 2.1 \\
$\text{HFe}$ & 6.5 & 2.9 & 7.6 & 3.0 & 3.0 & 1.1 & 4.9 & 2.5 \\
$\text{$\text{HF}\mu$}$ & 12.6 & 2.7 & 15.7 & 3.2 & 6.3 & 1.8 & 5.7 & 1.7 \\
$\text{Mat Conv}$ & 7.6 & 2.5 & 5.5 & 1.6 & 2.73 & 0.83 & 3.3 & 1.2 \\
$\text{Int Conv}$ & 2.7 & 1.6 & 3.0 & 1.7 & 2.1 & 1.2 & 2.7 & 1.6 \\
$\text{QMisID}$ & 8.1 & 2.2 & 8.1 & 2.2 & 1.48 & 0.39 & 1.48 & 0.39 \\
$\text{Other}$ & 20.3 & 5.5 & 13.3 & 4.0 & 9.3 & 2.8 & 7.0 & 2.7 \\
\midrule
Total Bkg. & 228 & 11 & 167.7 & 7.9 & 180 & 10 & 124.5 & 6.3 \\
\midrule
Data & \multicolumn{2}{c}{230} & \multicolumn{2}{c}{162} & \multicolumn{2}{c}{181} & \multicolumn{2}{c}{123} \\
\bottomrule
\end{tabular}
\end{table}

The fitted signal yield in each of the four SRs is found to be $\leq 0.001$ events and is therefore rounded to zero. A negative signal yield is not allowed in the fit as the quadratic EFT parameterization only allows for positive cross-sections by definition.
Since no significant signal is observed, upper limits on the three WCs are determined by running 1D- and 2D-likelihood scans.
Expected limits are also derived using a hybrid-Asimov data sample~\cite{Cowan:2010js}, which combines real data distributions in all nine CRs and the first two bins of each SRs, with Asimov data for the remaining bins determined under the background-only assumption.

The results of Ref. \cite{10.21468/SciPostPhysCore.6.1.013} indicate that Wilks' Theorem is violated by quadratic EFT terms, which can lead to over- or undercoverage for the derived confidence intervals (CIs). It is carefully checked if this affects the results of this analysis by comparing the observed limits with limits that are derived from MC pseudo-data samples. No undercoverage is observed, while overcoverage is observed for WC values close to 0. The observed limits are slightly affected by overcoverage leading to at most 9\% looser limits.

In the likelihood scans, the WCs are varied in the range of [$-0.01, 0.01$] for $c_{tu}^{(1)}$, [$-0.03, 0.03$] for $c_{Qu}^{(1)}$, and [$-0.06, 0.06$] for $c_{Qu}^{(8)}$. All NPs, normalization factors, and the other two WCs are allowed to vary freely. The resulting observed (expected) limits at 95\% CL in ($\TeV/\Lambda$)$^2$ are $|c_{tu}^{(1)}| < 0.0068\,(0.0071)$, $|c_{Qu}^{(1)}| < 0.020\,(0.022)$ and $|c_{Qu}^{(8)}| < 0.041\,(0.046)$.

These observed (expected) WCs limits are translated into 95\% CL cross-section limits on SS top-quark pair production by using the fitted EFT parameterization, varying only one WC at a time. For all three WCs the observed (expected) cross-section limit at 95\% CL is $1.6\,(2.0)$~fb.

In Table~\ref{tab:1D_Limits}, the observed 95\% CIs for the three WCs are shown for different sets of uncertainties, considering either only statistical uncertainties, or statistical and modeling uncertainties, or all uncertainties.
The sensitivity of the analysis is limited by the available number of events. A degradation of the observed limits of about 3\% is observed when including all systematic uncertainties, with the largest degradation caused by the modeling uncertainties for \ttW. The dominant uncertainties are indeed modeling uncertainties related to the ME and the PS of \ttW events.

In Figure~\ref{fig:1D_Limits_Compare}, the observed CIs on the three WCs are compared with the previous best constraints from the ATLAS Run~1 analysis of events with $b$-tagged jets and a SS lepton pair~\cite{EXOT-2013-16}. For the comparison with the ATLAS Run~1 analysis, the 95\% CL cross-section limits are converted to constraints on the WCs using the fitted EFT parameterization, as no specific WC limits were provided by the original analysis. The observed limits in this analysis are found to be an order of magnitude more stringent than the limits from the ATLAS Run~1 analysis.

In Figure~\ref{fig:Lambda_Limits}, the observed lower limits on $\Lambda$ for the WCs values 0.01, 1 and $4\pi^2$ are compared with the results of other analyses.

Two-dimensional-likelihood scans are performed for the three WCs, whose results are shown in Figure~\ref{fig:2D_Limits}.

\begin{table}[htbp]
\centering
\caption{Observed 95\% CL confidence intervals on the three Wilson coefficients for different sets of uncertainties. The first, second and third rows show the limits obtained by considering, respectively: only the statistical uncertainties, both the statistical and modeling uncertainties, and all uncertainties. The new-physics scale is set to $\Lambda = 1\TeV$.}
\begin{tabular}{lccc}
\toprule
\multicolumn{1}{c}{} & \multicolumn{3}{c}{Wilson Coefficient CIs at 95\% CL  $(\, \times 10^{-2}$)} \\
\cmidrule(lr){2-4}
Uncertainties & $c_{tu}^{(1)}$ & $c_{Qu}^{(1)}$ & $c_{Qu}^{(8)}$ \\
\midrule
Statistical uncertainty only & [-0.65, 0.65] & [-1.9, 1.9] & [-3.9, 3.9] \\					%
Statistical + modeling uncertainties & [-0.67, 0.67] & [-1.9, 1.9] & [-4.0, 4.0] \\			%
\midrule
Total uncertainty & [-0.68, 0.68] & [-2.0, 2.0] & [-4.1, 4.1] \\						%
\bottomrule
\end{tabular}
\label{tab:1D_Limits}
\end{table}

\begin{figure}[H]
\centering
\includegraphics[width=0.85\textwidth]{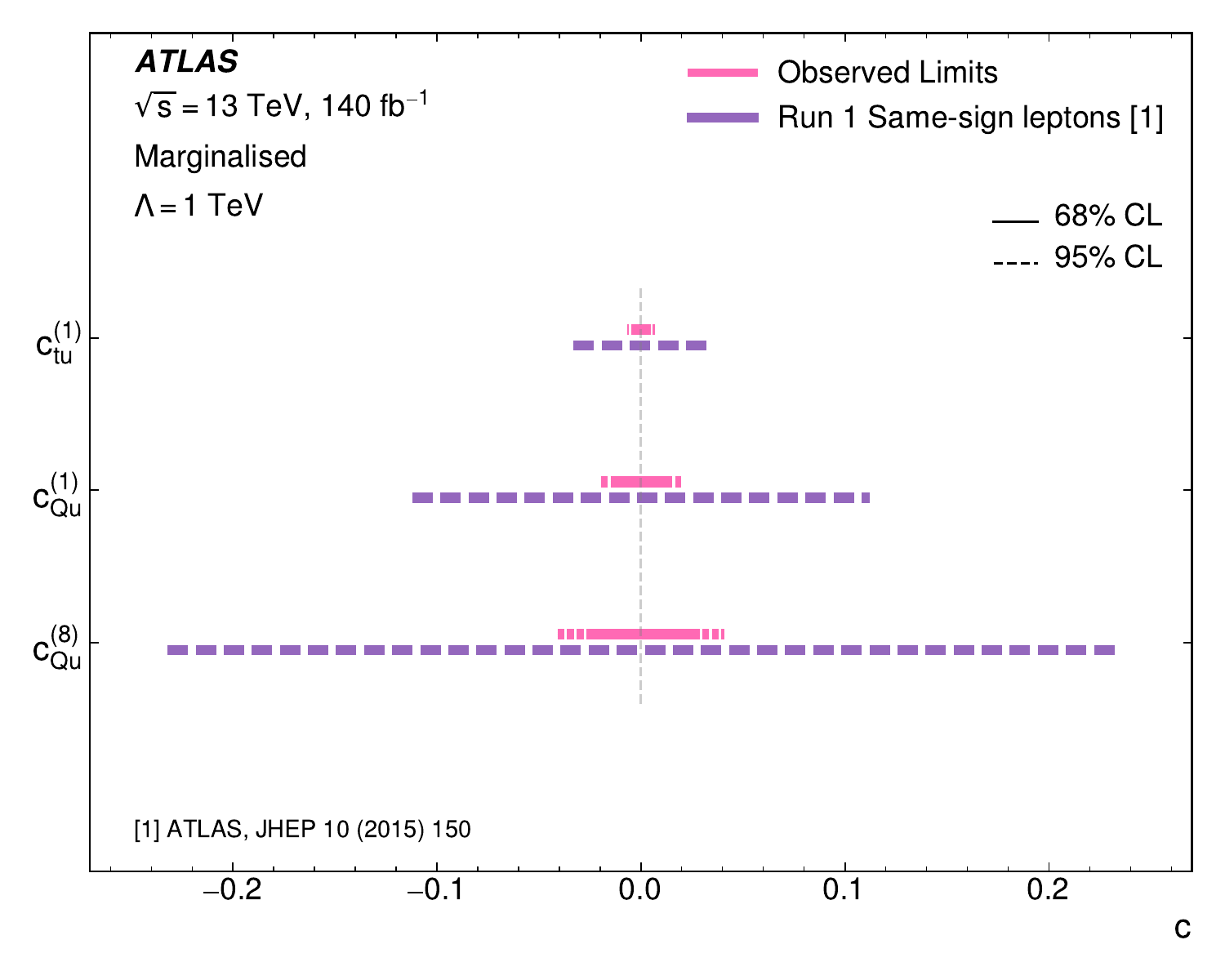}
\caption{Comparison of the observed limits on the three WCs to the limit obtained by the ATLAS Run~1 analysis of events with $b$-tagged jets and a same-sign lepton pair~\cite{EXOT-2013-16}. The cross-section limits from that analysis are converted to the limits on the WCs by using the fitted EFT parametrization. The bounds on the WCs are shown at the 68\% CL (solid) and/or 95\% CL (dashed) levels. For the ATLAS Run~1 analysis only the bounds at 95\% CL are available. The vertical bar represents the SM prediction. The new-physics scale is set to $\Lambda = 1\TeV$.}
\label{fig:1D_Limits_Compare}
\end{figure}

\begin{figure}[H]
\centering
\includegraphics[width=0.85\textwidth]{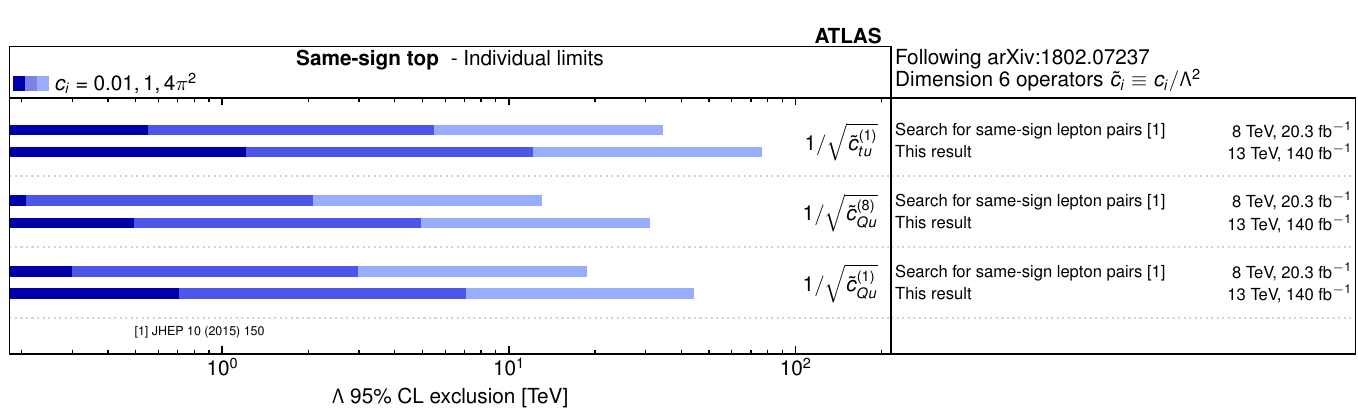}
\caption{Observed lower limits at 95\% confidence level on the scale of new physics $\Lambda$ for Wilson coefficient values of 0.01, 1 and $4\pi^2$. The limits are compared with the limits obtained by the ATLAS Run~1 analysis of events with $b$-tagged jets and a same-sign lepton pair~\cite{EXOT-2013-16}. \label{fig:Lambda_Limits}}
\end{figure}

\begin{figure}[H]
\centering
\subfloat[]{\includegraphics[width=0.49\textwidth]{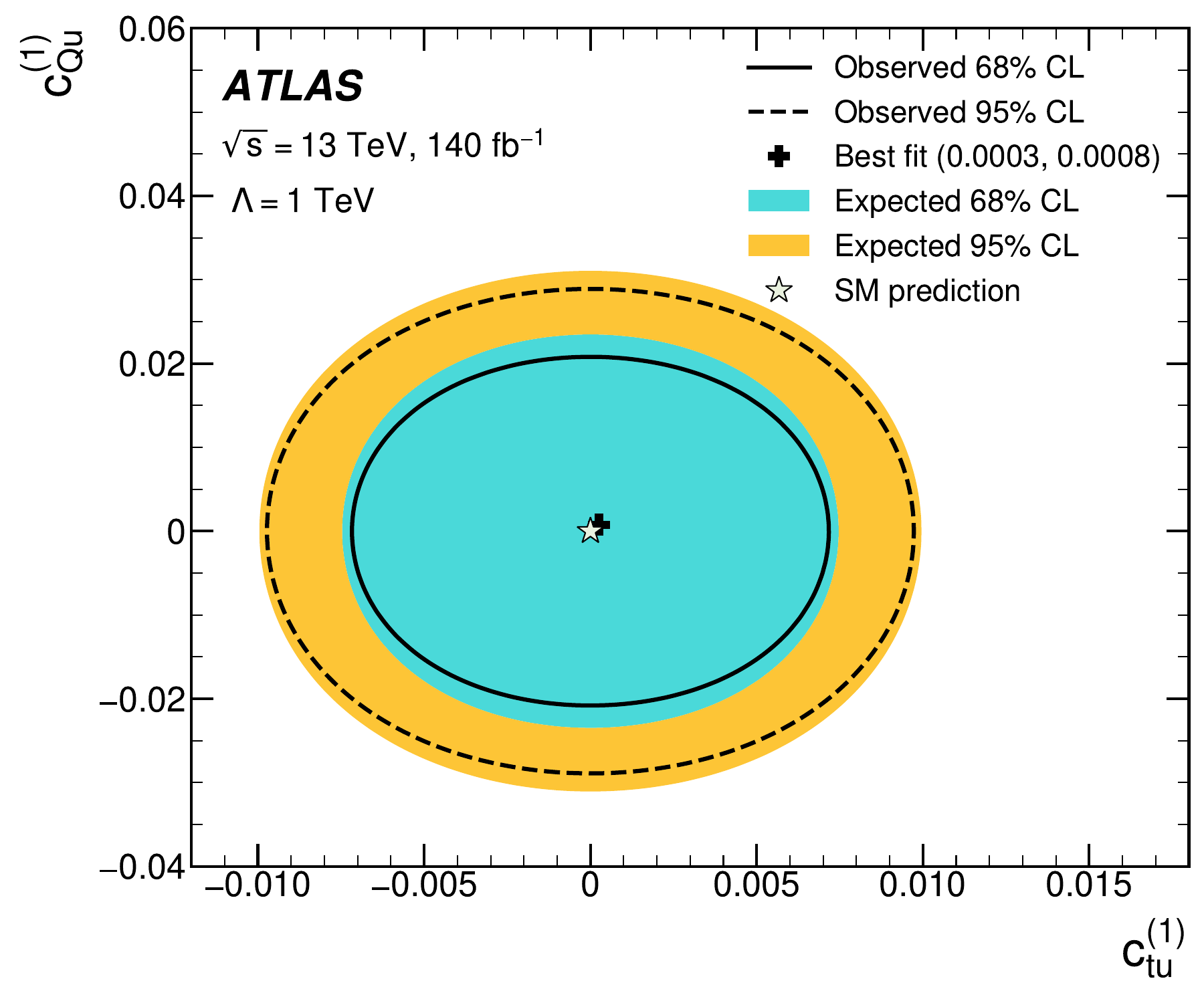}\label{fig:2Dcuu_cqu1}}
\subfloat[]{\includegraphics[width=0.49\textwidth]{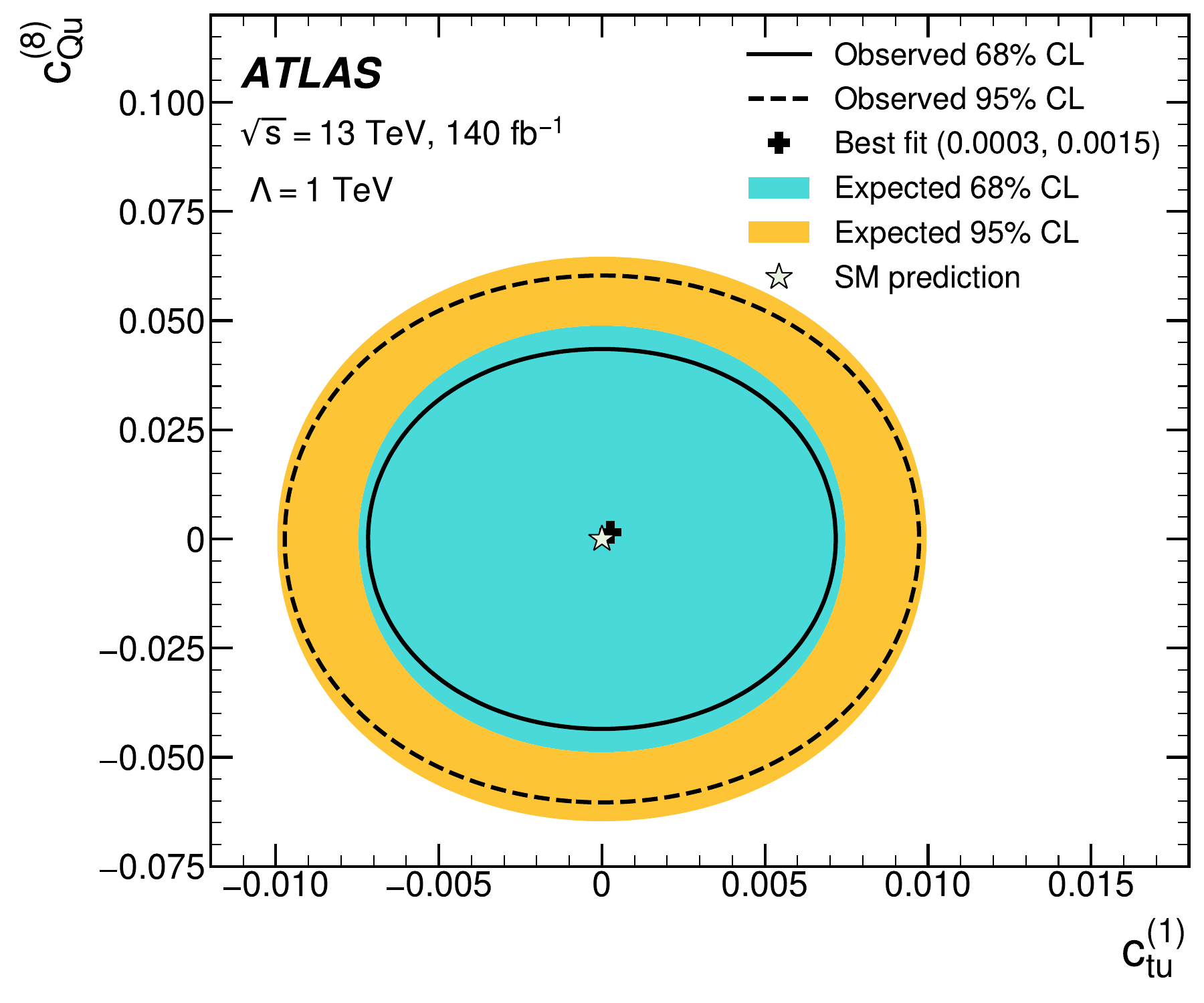}\label{fig:2Dcqu8_cuu}}\\
\subfloat[]{\includegraphics[width=0.49\textwidth]{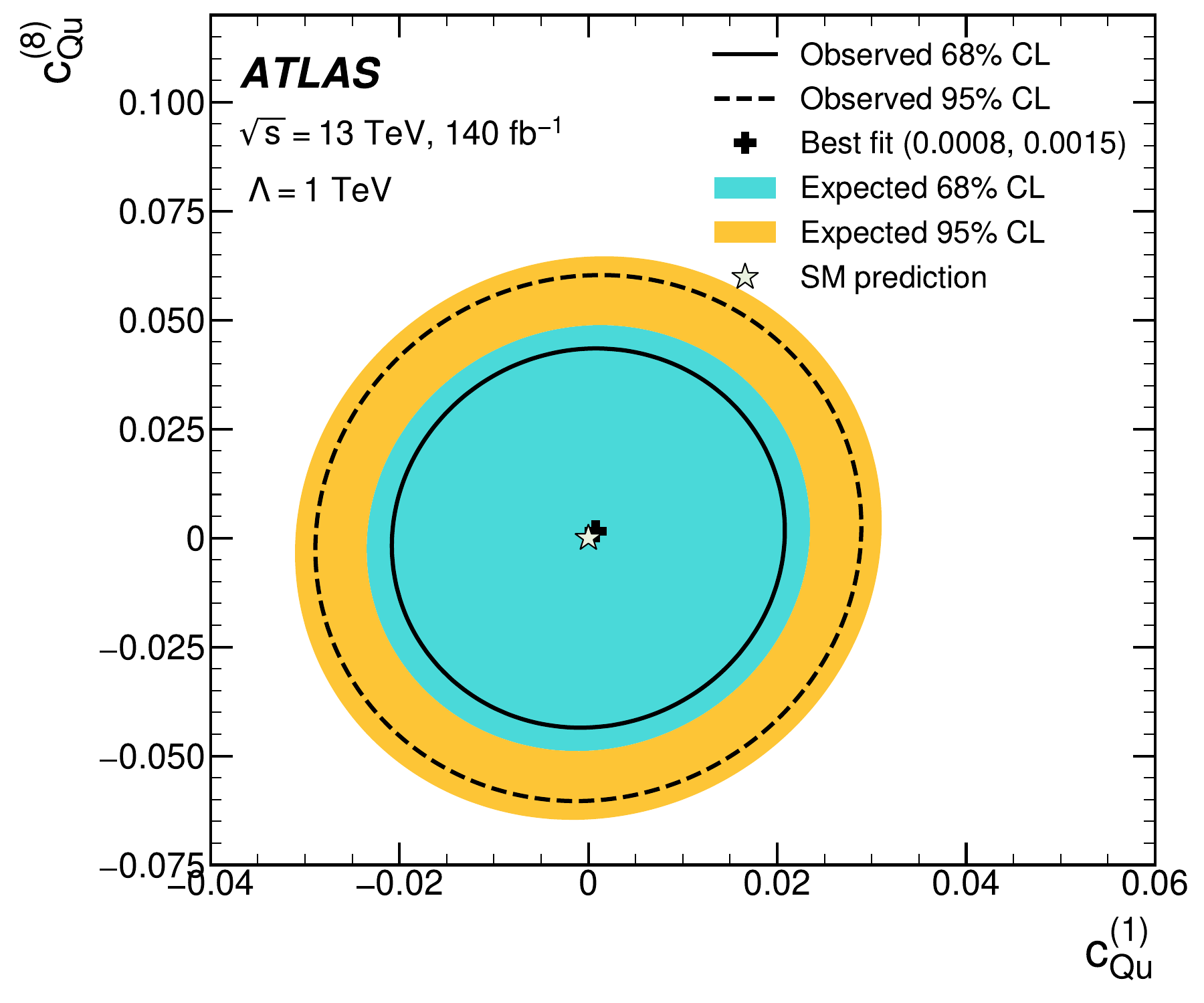}\label{fig:2Dcqu1_cqu8}}
\caption{2D likelihood scans for the different combinations of Wilson coefficients: (a) $c_{tu}^{(1)}$ versus $c_{Qu}^{(1)}$, (b) $c_{tu}^{(1)}$ versus $c_{Qu}^{(8)}$ and (c) $c_{Qu}^{(1)}$ versus $c_{Qu}^{(8)}$. The observed limits at 68\% and 95\% CL are represented by solid and dashed lines, respectively, while the expected limits are illustrated by the inner and outer shaded regions. The best fit value is marked with a cross. The SM prediction is marked with a star. The new-physics scale is set to $\Lambda = 1\TeV$.}
\label{fig:2D_Limits}
\end{figure}


\FloatBarrier

\section{Conclusion}
\label{sec:conclusion}

A search for the production of top-quark pairs with the same electric charge is reported. This search uses the full Run~2 data sample of $\sqrt{s}=13$~\TeV proton--proton collision data, with an integrated luminosity of 140~fb$^{-1}$, recorded from 2015 to 2018 with the ATLAS detector at the Large Hadron Collider. Standard model effective field theory is used to simulate the signal process, considering three Wilson coefficients associated with the ${\cal O}_{tu}^{(1)}$, ${\cal O}_{Qu}^{(1)}$, and ${\cal O}_{Qu}^{(8)}$ operators with the new physics scale set to $\Lambda = 1$~\TeV. Neural networks are employed to define signal regions sensitive to these Wilson coefficients. The largest background processes are constrained using dedicated control regions. The results are in agreement with the SM, with no significant signal detected. Upper limits at 95\% CL are determined for the three WCs and on the production cross-section of same-sign top-quark pairs. The observed (expected) limits on the WCs are $|c_{tu}^{(1)}| < 0.0068\,(0.0071)$, $|c_{Qu}^{(1)}| < 0.020\,(0.022)$ and $|c_{Qu}^{(8)}| < 0.041\,(0.046)$ respectively. The observed (expected) upper limit on the total production cross-section of same-sign top-quark pairs $\sigma(pp \rightarrow tt)$ is $1.6\,(2.0) $~fb at 95\% CL\@. These are the most stringent limits on the WCs associated with the ${\cal O}_{tu}^{(1)}$, ${\cal O}_{Qu}^{(1)}$, and ${\cal O}_{Qu}^{(8)}$ operators to date, improving previous limits by approximately a factor of 10.


%
\section*{Acknowledgements}
%

%
%

%
%

We thank CERN for the very successful operation of the LHC and its injectors, as well as the support staff at
CERN and at our institutions worldwide without whom ATLAS could not be operated efficiently.

The crucial computing support from all WLCG partners is acknowledged gratefully, in particular from CERN, the ATLAS Tier-1 facilities at TRIUMF/SFU (Canada), NDGF (Denmark, Norway, Sweden), CC-IN2P3 (France), KIT/GridKA (Germany), INFN-CNAF (Italy), NL-T1 (Netherlands), PIC (Spain), RAL (UK) and BNL (USA), the Tier-2 facilities worldwide and large non-WLCG resource providers. Major contributors of computing resources are listed in Ref.~\cite{ATL-SOFT-PUB-2023-001}.

We gratefully acknowledge the support of ANPCyT, Argentina; YerPhI, Armenia; ARC, Australia; BMWFW and FWF, Austria; ANAS, Azerbaijan; CNPq and FAPESP, Brazil; NSERC, NRC and CFI, Canada; CERN; ANID, Chile; CAS, MOST and NSFC, China; Minciencias, Colombia; MEYS CR, Czech Republic; DNRF and DNSRC, Denmark; IN2P3-CNRS and CEA-DRF/IRFU, France; SRNSFG, Georgia; BMBF, HGF and MPG, Germany; GSRI, Greece; RGC and Hong Kong SAR, China; ISF and Benoziyo Center, Israel; INFN, Italy; MEXT and JSPS, Japan; CNRST, Morocco; NWO, Netherlands; RCN, Norway; MNiSW, Poland; FCT, Portugal; MNE/IFA, Romania; MSTDI, Serbia; MSSR, Slovakia; ARIS and MVZI, Slovenia; DSI/NRF, South Africa; MICIU/AEI, Spain; SRC and Wallenberg Foundation, Sweden; SERI, SNSF and Cantons of Bern and Geneva, Switzerland; NSTC, Taipei; TENMAK, T\"urkiye; STFC/UKRI, United Kingdom; DOE and NSF, United States of America.

Individual groups and members have received support from BCKDF, CANARIE, CRC and DRAC, Canada; CERN-CZ, FORTE and PRIMUS, Czech Republic; COST, ERC, ERDF, Horizon 2020, ICSC-NextGenerationEU and Marie Sk{\l}odowska-Curie Actions, European Union; Investissements d'Avenir Labex, Investissements d'Avenir Idex and ANR, France; DFG and AvH Foundation, Germany; Herakleitos, Thales and Aristeia programmes co-financed by EU-ESF and the Greek NSRF, Greece; BSF-NSF and MINERVA, Israel; NCN and NAWA, Poland; La Caixa Banking Foundation, CERCA Programme Generalitat de Catalunya and PROMETEO and GenT Programmes Generalitat Valenciana, Spain; G\"{o}ran Gustafssons Stiftelse, Sweden; The Royal Society and Leverhulme Trust, United Kingdom.

In addition, individual members wish to acknowledge support from Armenia: Yerevan Physics Institute (FAPERJ); CERN: European Organization for Nuclear Research (CERN PJAS); Chile: Agencia Nacional de Investigaci\'on y Desarrollo (FONDECYT 1230812, FONDECYT 1230987, FONDECYT 1240864); China: Chinese Ministry of Science and Technology (MOST-2023YFA1605700, MOST-2023YFA1609300), National Natural Science Foundation of China (NSFC - 12175119, NSFC 12275265, NSFC-12075060); Czech Republic: Czech Science Foundation (GACR - 24-11373S), Ministry of Education Youth and Sports (FORTE CZ.02.01.01/00/22\_008/0004632), PRIMUS Research Programme (PRIMUS/21/SCI/017); EU: H2020 European Research Council (ERC - 101002463); European Union: European Research Council (ERC - 948254, ERC 101089007), European Union, Future Artificial Intelligence Research (FAIR-NextGenerationEU PE00000013), Italian Center for High Performance Computing, Big Data and Quantum Computing (ICSC, NextGenerationEU); France: Agence Nationale de la Recherche (ANR-20-CE31-0013, ANR-21-CE31-0013, ANR-21-CE31-0022, ANR-22-EDIR-0002); Germany: Baden-Württemberg Stiftung (BW Stiftung-Postdoc Eliteprogramme), Deutsche Forschungsgemeinschaft (DFG - 469666862, DFG - CR 312/5-2); Italy: Istituto Nazionale di Fisica Nucleare (ICSC, NextGenerationEU), Ministero dell'Università e della Ricerca (PRIN - 20223N7F8K - PNRR M4.C2.1.1); Japan: Japan Society for the Promotion of Science (JSPS KAKENHI JP22H01227, JSPS KAKENHI JP22H04944, JSPS KAKENHI JP22KK0227, JSPS KAKENHI JP23KK0245); Netherlands: Netherlands Organisation for Scientific Research (NWO Veni 2020 - VI.Veni.202.179); Norway: Research Council of Norway (RCN-314472); Poland: Ministry of Science and Higher Education (IDUB AGH, POB8, D4 no 9722), Polish National Agency for Academic Exchange (PPN/PPO/2020/1/00002/U/00001), Polish National Science Centre (NCN 2021/42/E/ST2/00350, NCN OPUS 2023/51/B/ST2/02507, NCN OPUS nr 2022/47/B/ST2/03059, NCN UMO-2019/34/E/ST2/00393, NCN \& H2020 MSCA 945339, UMO-2020/37/B/ST2/01043, UMO-2021/40/C/ST2/00187, UMO-2022/47/O/ST2/00148, UMO-2023/49/B/ST2/04085, UMO-2023/51/B/ST2/00920); Slovenia: Slovenian Research Agency (ARIS grant J1-3010); Spain: Generalitat Valenciana (Artemisa, FEDER, IDIFEDER/2018/048), Ministry of Science and Innovation (MCIN \& NextGenEU PCI2022-135018-2, MICIN \& FEDER PID2021-125273NB, RYC2019-028510-I, RYC2020-030254-I, RYC2021-031273-I, RYC2022-038164-I); Sweden: Carl Trygger Foundation (Carl Trygger Foundation CTS 22:2312), Swedish Research Council (Swedish Research Council 2023-04654, VR 2018-00482, VR 2022-03845, VR 2022-04683, VR 2023-03403, VR grant 2021-03651), Knut and Alice Wallenberg Foundation (KAW 2018.0458, KAW 2019.0447, KAW 2022.0358); Switzerland: Swiss National Science Foundation (SNSF - PCEFP2\_194658); United Kingdom: Leverhulme Trust (Leverhulme Trust RPG-2020-004), Royal Society (NIF-R1-231091); United States of America: U.S. Department of Energy (ECA DE-AC02-76SF00515), Neubauer Family Foundation.

%
%


%
\clearpage
\printbibliography
\clearpage
 
\begin{flushleft}
\hypersetup{urlcolor=black}
{\Large The ATLAS Collaboration}

\bigskip

\AtlasOrcid[0000-0002-6665-4934]{G.~Aad}$^\textrm{\scriptsize 105}$,
\AtlasOrcid[0000-0001-7616-1554]{E.~Aakvaag}$^\textrm{\scriptsize 17}$,
\AtlasOrcid[0000-0002-5888-2734]{B.~Abbott}$^\textrm{\scriptsize 124}$,
\AtlasOrcid[0000-0002-0287-5869]{S.~Abdelhameed}$^\textrm{\scriptsize 120a}$,
\AtlasOrcid[0000-0002-1002-1652]{K.~Abeling}$^\textrm{\scriptsize 57}$,
\AtlasOrcid[0000-0001-5763-2760]{N.J.~Abicht}$^\textrm{\scriptsize 51}$,
\AtlasOrcid[0000-0002-8496-9294]{S.H.~Abidi}$^\textrm{\scriptsize 30}$,
\AtlasOrcid[0009-0003-6578-220X]{M.~Aboelela}$^\textrm{\scriptsize 46}$,
\AtlasOrcid[0000-0002-9987-2292]{A.~Aboulhorma}$^\textrm{\scriptsize 36e}$,
\AtlasOrcid[0000-0001-5329-6640]{H.~Abramowicz}$^\textrm{\scriptsize 156}$
% ...
\bigskip
\\

% ...
$^{17}$Department for Physics and Technology, University of Bergen, Bergen; Norway.\\
% ...
$^{30}$Physics Department, Brookhaven National Laboratory, Upton NY; United States of America.\\
% ...
$^{36}$$^{(a)}$Facult\'e des Sciences Ain Chock, Universit\'e Hassan II de Casablanca;$^{(b)}$Facult\'{e} des Sciences, Universit\'{e} Ibn-Tofail, K\'{e}nitra;$^{(c)}$Facult\'e des Sciences Semlalia, Universit\'e Cadi Ayyad, LPHEA-Marrakech;$^{(d)}$LPMR, Facult\'e des Sciences, Universit\'e Mohamed Premier, Oujda;$^{(e)}$Facult\'e des sciences, Universit\'e Mohammed V, Rabat;$^{(f)}$Institute of Applied Physics, Mohammed VI Polytechnic University, Ben Guerir; Morocco.\\
% ...
$^{46}$Physics Department, Southern Methodist University, Dallas TX; United States of America.\\
% ...
$^{51}$Fakult\"{a}t Physik , Technische Universit{\"a}t Dortmund, Dortmund; Germany.\\
% ...
$^{57}$II. Physikalisches Institut, Georg-August-Universit\"{a}t G\"ottingen, G\"ottingen; Germany.\\
% ...
$^{105}$CPPM, Aix-Marseille Universit\'e, CNRS/IN2P3, Marseille; France.\\
% ...
$^{120}$$^{(a)}$New York University Abu Dhabi, Abu Dhabi;$^{(b)}$United Arab Emirates University, Al Ain; United Arab Emirates.\\
% ...
$^{124}$Homer L. Dodge Department of Physics and Astronomy, University of Oklahoma, Norman OK; United States of America.\\
% ...
$^{156}$Raymond and Beverly Sackler School of Physics and Astronomy, Tel Aviv University, Tel Aviv; Israel.\\
% ...
$^{a}$ Also Affiliated with an institute covered by a cooperation agreement with CERN.\\
$^{b}$ Also at An-Najah National University, Nablus; Palestine.\\
$^{c}$ Also at Borough of Manhattan Community College, City University of New York, New York NY; United States of America.\\
$^{d}$ Also at Center for High Energy Physics, Peking University; China.\\
$^{e}$ Also at Center for Interdisciplinary Research and Innovation (CIRI-AUTH), Thessaloniki; Greece.\\
$^{f}$ Also at CERN, Geneva; Switzerland.\\
% ...

\end{flushleft}


%

%
%

%

%
%
%
%

\end{document}